\documentclass[11pt]{article}
\usepackage[pdfborder={0 0 0 0}]{hyperref}

\usepackage{mathrsfs}
\usepackage{amsfonts}
\usepackage{amsmath}
\usepackage{amssymb}
\usepackage[english]{babel}
\usepackage{feynmf}
\usepackage{simplewick}
\usepackage{graphicx}
\usepackage{float}
\usepackage{fancyhdr}
\usepackage{slashed}
\usepackage{xspace}
\usepackage{multirow}
\usepackage{tabularx}

\def\be{\begin{equation}}
\def\ee{\end{equation}}
\def\bea{\begin{eqnarray}}
\def\eea{\end{eqnarray}}

\def\cal{\mathcal}

\global \long \def \l{\lambda}

\global \long \def \r{\rho}

\global \long \def \Jm{\mathcal{J}}

\global \long \def \Tm{\mathcal{T}}

\global \long \def \D{\Delta}
\global \long \def \a{\alpha}

\global \long \def \eps{\epsilon}
\global \long \def \veps{\varepsilon}

\global \long \def \pd{\partial}

\newcommand{\op}{\ensuremath{\mathcal{O}}\xspace}
\newcommand{\del}{\partial} 

\newcommand{\vev}[1]{\ensuremath{\langle #1 \rangle}\xspace}

\def\ie{{\it i.e.\ }}

\newcommand{\hf}{\frac{1}{2}}
\newcommand{\tf}{\frac{3}{2}}
\newcommand{\qt}{\frac{1}{4}}

\let\a=\alpha    \let\e=\epsilon
    
\let\l=\lambda \let\m=\mu \let\n=\nu  \let\r=\rho
\let\s=\sigma     
  \let\D=\Delta  
    
    \let\G=\Gamma

\usepackage{fullpage}

\numberwithin{equation}{section}

\date{\today}

\begin{document}

\begin{titlepage}

\vbox{\hfill YITP-SB-12-37}
\vbox{}
\vbox{}
\vbox{}
\begin{center}
{\bf\LARGE{The Bootstrap Program for Boundary CFT$_d$}}\vskip2.25cm{
Pedro Liendo\footnote{\texttt{pedro.liendo@stonybrook.edu}},
Leonardo Rastelli\footnote{\texttt{leonardo.rastelli@stonybrook.edu}},
Balt C. van Rees\footnote{\texttt{vanrees@insti.physics.sunysb.edu}}
}\vskip1cm 
{\textit{C.~N.~Yang Institute for Theoretical Physics, Stony Brook University}\\
{\textit{Stony Brook, NY 11794-3840}}}
\end{center}
\vskip1.25cm

\begin{abstract}
\vskip.15cm
\noindent
We study the constraints of crossing symmetry and unitarity for conformal field theories in the presence of a boundary, 
with a focus on the Ising model in various dimensions. We show that an analytic approach to the bootstrap is feasible for free-field theory and at one loop in the epsilon expansion, but more generally one has to resort to numerical methods. Using the recently developed linear programming techniques we find several interesting bounds for operator dimensions and OPE coefficients and comment on their physical relevance. We also show that the ``boundary bootstrap'' can be  easily applied to correlation functions of tensorial operators and study the stress tensor as an example. In the appendices we present conformal block decompositions of a variety of physically interesting correlation functions.
\end{abstract}
\vfill
\today
\end{titlepage}

\tableofcontents

\section{Introduction}

The ``bootstrap''  has been a recurring dream in theoretical physics. It is
  the ambitious aspiration that, starting from a few basic spectral assumptions, 
  symmetries and 
 general consistency requirements (such as unitarity and crossing)
will be powerful enough to  fix  the form of the theory, with no reference to a Lagrangian.
The dual models of the strong interactions
emerged as an incarnation of the $\mathrm{S}$-matrix bootstrap attempts of the 1960s and eventually led to the discovery of string theory.
The bootstrap program  for conformal field theories (CFTs) in $d$ dimensions was formulated in the early 1970s \cite{Ferrara:1973yt,Ferrara:1973vz,Polyakov:1974gs}. 
Despite important formal developments such as the operator product expansion and the conformal block decomposition
(see {\it e.g.} the early books \cite{ferrarabook,todorovbook}),  
attempts to solve CFTs in arbitrary dimensions
were not successful. For {two-dimensional} CFTs, the revolution came in the 1980s with the discovery of many exactly-solvable  ``rational'' models. 
While  this is a beautiful incarnation of the bootstrap idea, the methods that work
 in $2d$ rational CFTs\footnote{or in closely-related models such as Liouville theory} are too specialized
to be imitated in higher dimensions, or even in two dimensions for the generic non-rational model.

The interest in CFT in various dimensions is nowadays stronger than ever, sustained
by  phenomenological questions in condensed matter physics ($d=3$) and particle physics ($d=4$),
as well as by more formal motivations such as the AdS/CFT correspondence and the rich integrability
structures of  superconformal field theories ($d \leq 6$).
A pioneering work \cite{Rattazzi:2008pe} has rekindled the conformal bootstrap,
turning it into a concrete computational tool.
This approach has been refined and extended in a series of papers \cite{Rychkov:2009ij, Caracciolo:2009bx, Rattazzi:2010gj,Poland:2010wg,Rattazzi:2010yc,Vichi:2011ux,Poland:2011ey, Rychkov:2011et, ElShowk:2012ht}. 

The modern bootstrap starts with the simple question:
in a generic theory, which   values of operator dimensions
 and  OPE coefficients   are compatible with the constraints of crossing symmetry and unitarity for the four-point functions?
 There is a shift of viewpoint, from trying to find analytic answers in a specific model to deriving  (by numerical methods if necessary)
 universal bounds valid for any model.
 As it turns out, one can derive strong constraints already from
 the analysis of a single four-point function of identical scalar operators \cite{Rattazzi:2008pe}.  This should be regarded as the first step in a systematic exploration of the space of CFTs.
 More surprisingly, important theories such as the $3d$ Ising model appear to live at interesting corners of the parameter space,
 sitting at ``kinks'' of the exclusion curves  \cite{Rychkov:2009ij,Rychkov:2011et, ElShowk:2012ht}. So even the  solution of some special models in $d >2$
 may not be too far-fetched, after all.

In its simplest version, the revived conformal bootstrap works as follows. The four-point correlation function $\vev{\varphi(x_1) \varphi(x_2) \varphi(x_3) \varphi(x_4)}$ of a scalar operator can
be written as a sum over conformal blocks in two different channels, by taking OPEs in two different limits.
The 
conformal block decompositions in either channel must sum to the same four-point function,  giving crossing-symmetry relations for the couplings and scaling dimensions.
While this was understood long ago, the main idea of \cite{Rattazzi:2008pe} is that these constraints can be put to good use by taking derivatives of the four-point function at symmetric points
and applying linear programming techniques  to obtain contradictions if certain conditions for {\it e.g.}  the operator spectrum are not met. The prototypical example of a constraint that arises in this way is an upper bound for the dimension of the first  scalar primary $\varphi^2$ appearing in the OPE of two $\varphi$'s.
Crossing symmetry and unitarity imply that 
$\Delta_{\varphi^2}  \leq f(\Delta_\varphi)$ for some numerically determined function $f(\Delta_\varphi)$. The method admits straightforward extensions to  bounds on scaling dimensions of  tensorial operators, central charges and OPE coefficients.\footnote{Analogous ``sum rule'' techniques can also be used to obtain non-trivial bounds from modular invariant partition functions, see \cite{Cardy:1986ie,Cardy:1991kr,Hellerman:2009bu,Hellerman:2010qd,Friedan:2012jk}.}

In this paper we extend this program to conformal field theories with a boundary.
An Euclidean CFT  in $d$ dimensions can be defined in the half-space $x_d \geq 0$,
with boundary conditions at $x_d =0$ that preserve an $SO(d,1)$ subgroup of the original $SO(d+1,1)$ conformal symmetry \cite{Cardy:1984bb,binderbook8}. 
For a given bulk CFT, different consistent boundary conditions are usually possible.
Boundary CFTs (BCFTs) are very interesting in their own right and find diverse physical applications.
They describe surface phenomena in systems
near criticality, 
with surface critical exponents related to the conformal dimensions of the boundary operators.
 In string theory, two-dimensional  worldsheet BCFTs are interpreted as D-branes.  
 These would be sufficient reasons to consider the boundary bootstrap, 
 but one of the main questions we would like to address is
 whether by probing the theory
 with a boundary one can constrain the original bulk theory itself.\footnote{A prototype is the
the beautiful theory developed by Cardy \cite{Cardy:1989ir,Cardy:1991tv} in $2d$ rational CFTs, which relates the set of consistent boundary conditions
 with the bulk spectrum and its modular transformation properties.}
 One could in fact also go ahead and consider a more general setup where conformal defects of all possible codimensions
  (boundaries being the special case  of codimension one) appear on a democratic footing.

 Besides the spectrum of bulk operators and their three-point functions,
which are unaffected by the boundary conditions, a BCFT  is characterized by additional boundary data:
the spectrum of boundary operators, their three-point functions, and the bulk-boundary two-point functions.
A correlator containing both bulk and boundary operators can be decomposed in different channels, giving crossing-symmetry constraints that in general 
involve both bulk and boundary data.
We will focus on
the simplest non-trivial type of correlator, the two-point function of two
bulk operators,
which in the presence of a boundary is a non-trivial function of a {\it single} conformal cross-ratio.  
It can be decomposed in the bulk channel, by first fusing the two bulk operators together,
or in the boundary channel, by taking the boundary OPE of each bulk operator. See figure \ref{csfigure} on page \pageref{csfigure}.

\subsection*{Outline}
The main advantage of using the boundary bootstrap to constrain bulk dynamics is the simplicity of the setup just described. This follows from the results of section \ref{sec:bdycrossingsym}, where we discuss the two-point function of bulk scalar operators: its functional form and its conformal block decomposition in the bulk and boundary channels. The conformal blocks turn out to be simple (hypergeometric) functions of the single cross-ratio and furthermore depend analytically on the spacetime dimension $d$. This is to be contrasted with the standard conformal blocks for four-point functions (in a theory with no boundary), which depend on two cross-ratios and admit closed-form expressions only when $d$ is an even integer.

In section \ref{sec:bootstrapinepsexpansion} we demonstrate a remarkable simplification of the boundary bootstrap in a few special cases, where one can explicitly solve the bootstrap equations by making an ansatz containing only a few conformal blocks in either channel. By this route we are able to recover one-loop results in the epsilon expansion purely from the bootstrap equations. These 
 methods do not straightforwardly  extend to higher loops but give a nice pedagogical illustration of the constraining power of crossing symmetry.

In section \ref{sec:scalarnumerics} we apply the linear programming techniques of \cite{Rattazzi:2008pe} to the boundary crossing symmetry equations for  scalar two-point functions. We derive a number of general bounds on operator dimensions and OPE coefficients. Our bounds however come with a major caveat: while unitarity
guarantees that the coefficients in the boundary conformal block expansion are  positive (since they are squares of real numbers, as in \cite{Rattazzi:2008pe}),
this is not automatically the case for the bulk expansion. Indeed it is not difficult to find counterexamples
for certain choices of boundary conditions.
We  need then to  \emph{assume} the existence of boundary conditions where
 the coefficients multiplying the bulk conformal blocks are positive.
 We present circumstantial evidence for this  assumption in the appendices
 where we show that it holds in a large number of calculable cases, for favorable choices of the boundary conditions (the so-called ``extraordinary'' and  ``special'' transitions).
  It would however be more satisfactory to find a general proof.

External tensorial bulk operators are also more easily incorporated in the boundary setup. We illustrate this in the second half of the paper,
where we consider
  the two-point function of two bulk stress tensors. In section \ref{sec:crossingsyenmom} we discuss the different tensorial structures, the bulk and boundary conformal block decompositions and the resulting crossing symmetry equations. We then apply the linear programming techniques in section \ref{sec:Tmnbounds} and derive interesting
   bounds.
As before, these results  rely on certain positivity assumptions for the coefficients of the bulk conformal blocks.

In appendix \ref{app:scalarblockders} we present a brief derivation of the conformal blocks for a scalar two-point function. The remaining two appendices are dedicated to a discussion of a large number of solutions to the crossing symmetry equations: we consider scalar two-point functions in appendix \ref{app:crossingsysolns} and stress-tensor two-point functions   in appendix \ref{app:enmomtencorrs}. These solutions offer partial justification of our positivity assumptions in sections \ref{sec:scalarnumerics} and \ref{sec:Tmnbounds}. We also consider an interesting two-point function in Liouville theory (with ZZ boundary conditions)
that interpolates between all  the minimal models. We discuss how the analogous
bulk four-point function helps to explain a few features of the ``kinks'' observed in the bulk results of \cite{Rychkov:2009ij,Rychkov:2011et, ElShowk:2012ht}.

\section{Boundary crossing symmetry for scalars}
\label{sec:bdycrossingsym}

 In this section we introduce the general setup of boundary CFT and 
 derive the crossing symmetry 
equations for the two-point function of bulk scalar operators.
 For background material on BCFTs see \cite{Cardy:1984bb,Cardy:1991tv,Diehl:1981zz,McAvity:1993ue,DiFrancesco:1997nk}, and especially the paper by McAvity and Osborn \cite{McAvity:1995zd}, 
whose results we borrow at several points in this and subsequent sections. 

\subsection{Scalar two-point function}
\label{subsec:scalar2ptfunc}

Let us start by deriving the form of the scalar two-point function in the presence of a boundary, a classic result dating back 
to \cite{Cardy:1984bb}. We will use standard Euclidean coordinates $x^\mu = (x^1, \ldots , x^d)$ and consider the half-space defined by $x^d > 0$, the coordinates tangential to the boundary are denoted 
$\vec x$. It will be useful to embed this physical space in a higher dimensional space as the so-called null projective cone \cite{Dirac:1936fq,Mack:1969rr}. Consider Minkowski space in $d$+2 dimensions in lightcone coordinates denoted by $P^A = (P^+, P^-, P^1, \ldots P^d)$. The null projective cone is defined as,
\be
\label{cone}
P^A P_A=0 \hspace{1cm} \textrm{with} \hspace{1cm} P^A \sim \l P^A\,.
\ee
The map from the null projective cone to our physical space is given by
\be
x^\mu =  \frac{P^\mu}{P^+}\,.
\ee
One easily finds that the usual $SO(d+1,1)$ Lorentz group of the $d$+2-dimensional Minkowski space becomes the conformal group of the $d$-dimensional Euclidean space. The null projective cone 
provides a linearization of the action of the conformal group.

As we mentioned above, the presence of a boundary at $x^d = 0$ breaks the symmetry group to $SO(d,1)$. In the null projective cone this breaking can be implemented by introducting a fixed vector $V$ with components 
\be 
V^A = (0,\ldots,0,1)\, ,
\ee
and restricting ourselves to those Lorentz transformations that leave $V^A$ invariant. The residual conformal transformations for the coordinates $x^\mu$ are easily obtained from the linear transformations of the $P^A$ coordinates.

Let us now consider scalar fields that are homogeneous functions of the coordinates,
\be 
\label{homog}
\cal O(\l P)=\l^{-\Delta} \cal O(P)\, ,
\ee
where $\Delta$ is the conformal dimension of the field $\cal O$.
The physical CFT scalar operator is defined as
\be 
O(x) = (P^+)^\Delta \cal O(P)\, .
\ee
The two-point function of $\op$ should be invariant under $SO(d,1)$ and 
consistent with (\ref{homog}). The only $SO(d,1)$ invariants that can be formed with two coordinates and the fixed vector $V^A$ are
\be 
P_1 \cdot P_2, \hspace{0.5cm} V \cdot P_1, \hspace{0.5cm} \textrm{and} \hspace{0.5cm}V \cdot P_2 .
\ee 
The two-point function must then be of the form
\be 
\langle \cal O_1(P_1) \cal O_2(P_2) \rangle = \frac{1}{(2V \cdot P_1)^{\Delta_1}(2V \cdot P_2)^{\Delta_2}}f(\xi),
\ee
where $f(\xi)$ is an arbitrary function of the conformal invariant,
\be 
\xi = \frac{-P_1 \cdot P_2}{2(V \cdot P_1)(V \cdot P_2)}.
\ee
In physical coordinates,
\be 
\xi =\frac{(x_1-x_2)^2}{4x_1^d x_2^d}\,.
\ee
We see that the limit $\xi \to 0$ corresponds to bringing the operators close together while the limit $\xi \to \infty$ amounts to bringing the operators close to the boundary. It will be useful to introduce a function $G(\xi) = \xi^{(\Delta_1 + \Delta_2)/2} f(\xi)$, the two-point function then becomes
\be 
\label{twoptscalar}
\langle O_1(x_1) O_2(x_2) \rangle = \frac{1}{(2x_1^d)^{\Delta_1}(2x_2^d)^{\Delta_2}} \xi^{- (\Delta_1 + \Delta_2)/2} G(\xi).
\ee
For two identical (canonically normalized) operators $\lim_{\xi \to 0} G(\xi) = 1$, since we need to recover the usual two-point function far away from the boundary.
Although using the null projective cone  is somewhat of an overkill for the scalar two-point function, this formalism will become essential for the tensor calculations of section \ref{sec:crossingsyenmom}. 

\subsection{The boundary bootstrap}
\label{subsec:crossingsy}

Much like a four-point function for a CFT without a boundary, one can decompose the correlation function \eqref{twoptscalar} into conformal blocks. In this case there exist two different decompositions (or channels) 
and we review both of them below.

In the \emph{bulk channel} we simply substitute the bulk OPE in the two-point function \eqref{twoptscalar}. For two identical scalar operators the bulk OPE takes the form (omitting tensor indices for simplicity):
\be
\label{bulkOPEscalar}
O(x) O(y) = \frac{1}{(x-y)^{2\Delta}} + \sum_k \lambda_k C[x-y, \del_y] O_k(y)\, ,
\ee 
where $k$ labels conformal primary fields. The differential operators $C[x-y, \del_y]$ are determined by the (bulk) conformal symmetry and the couplings $\lambda_k$ can be taken to be real \cite{Rattazzi:2008pe}. We emphasize that this OPE is a local property of the bulk CFT and therefore unaffected by the presence of a boundary. On the other hand, whereas in the absence of any boundaries only the identity operator gets a non-zero one-point function (and all other terms in the OPE therefore drop out of the two-point function of $O$), this is no longer the case once a boundary is present. Using the null projective cone it is easily demonstrated that boundary conformal invariance allows for one-point functions of scalar operators of the form:
\be
\label{oneptscalar}
\vev{O(x)} = \frac{a_{O}}{(2 x^d)^\Delta}\, ,
\ee  
with a coefficient $a_{O}$ whose magnitude is unambiguous as we have normalized the operator using the first term in \eqref{bulkOPEscalar}. One-point functions for operators with spin are not allowed by conformal invariance, see section \ref{subsec:corrtensor} below. Substituting now \eqref{bulkOPEscalar} in \eqref{twoptscalar} and using \eqref{oneptscalar} one arrives at the bulk channel conformal block decomposition:
\be
\label{bulkblockdec}
G(\xi) = 1 + \sum_{k} \lambda_k a_{k} \,f_{\text{bulk}}(\Delta_k;\xi)\, ,
\ee
where the \emph{bulk conformal blocks} $f_{\text{bulk}}(\Delta_k;\xi)$ can be determined by working out the expression:
\be
C[x-y, \del_y] \frac{1}{(y^d)^{\Delta_k}}\, .
\ee
This computation was performed in \cite{McAvity:1995zd}, with the result that (see appendix \ref{app:scalarblockders} for a new derivation)
\be
\label{bulkblock}
f_{\text{bulk}}(\Delta_k;\xi) = \xi^{\Delta_k / 2} {}_2 F_1\left(\frac{\D_k}{2},\frac{\D_k}{2};\D_k + 1 - \frac{d}{2};-\xi\right)\, .
\ee
Equations \eqref{bulkblockdec} with the explicit expression \eqref{bulkblock} summarize the bulk block decomposition of the two-point function. Notice that the blocks are naturally defined as a series expansion around $\xi = 0$, which is when the two operators approach each other. Convergence of the OPE away from the boundary however implies that the conformal block decomposition should 
converge for all physical values of $\xi$, that is for all $0 < \xi < \infty$.

In the \emph{boundary channel} we use the bulk-to-boundary OPE where a bulk operator is written as an infinite sum over boundary operators. For a scalar operator this OPE takes the form:
\be
\label{bdyOPEscalar}
O(x) = \frac{a_O}{(2x^d)^\Delta} + \sum_l \mu_l D[x^d,\del_{\vec x}] \hat O_l(\vec x)\,,
\ee
where the index $l$ runs over boundary primary fields, the differential operators $D[x^d,\del_{\vec x}]$ are again completely determined by (boundary) conformal symmetry and the couplings $\mu_l$ are again assumed to be real. The first term in \eqref{bdyOPEscalar} corresponds to the one-point function of $O (x)$ and represents the contribution of the boundary identity operator. Subsequent operators all have to be scalars by boundary Lorentz invariance. Notice also that in equation \eqref{bdyOPEscalar} we used a hat to denote operators living on the boundary (and such operators obviously can depend only on $\vec x$). 

The constraints of boundary conformal invariance for the correlation functions of boundary operators $\hat O(\vec x)$ are exactly the same as those of ordinary conformal invariance in $d-1$ dimensions. This implies in particular that boundary operators cannot get one-point functions and their two-point functions take the canonical form,
\be
\label{twoptbdyscalar}
\vev{\hat O(\vec x) \hat O(\vec y)} = \frac{1}{|\vec x - \vec y|^{2\Delta}}\, ,
\ee
which also provides a normalization for boundary operators. Combining now \eqref{bdyOPEscalar} and \eqref{twoptscalar} and using \eqref{twoptbdyscalar} one arrives at the boundary channel conformal block decomposition:
\be
\label{boundaryblockdec}
G(\xi) = \xi^\Delta \left( a_O^2 + \sum_{l} \mu_l^2 \,f_{\text{bdy}}(\Delta_l;\xi) \right)\, ,
\ee
where the \emph{boundary conformal blocks} $f_{\text{bdy}}(\Delta_l;\xi)$ can now be determined from:
\be
 D[x^d,\del_{\vec x}]  D[y^d,\del_{\vec y}] \frac{1}{|\vec x - \vec y|^{2\Delta}}\,.
\ee
Just as for the bulk blocks, this computation was done in \cite{McAvity:1995zd} (and rederived in appendix \ref{app:scalarblockders}), 
\be
\label{boundaryblock}
f_{\text{bdy}}(\Delta;\xi) = \xi^{- \Delta} {}_2 F_1 \left(\Delta,\D + 1 - \frac{d}{2}; 2 \D + 2 - d; - \frac{1}{\xi}\right)\, .
\ee
The boundary blocks have a good series expansion when both operators approach the boundary, that is around $\xi = \infty$.

The boundary block decomposition is summarized by equations \eqref{boundaryblockdec} and \eqref{boundaryblock}. The convergence of the bulk-boundary OPE away from other operator insertions implies that this conformal block decomposition should converge for all $0 < \xi < \infty$ as well.

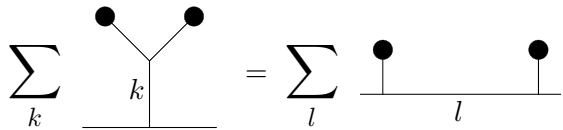
\begin{figure}[H]
             \begin{center}           
              \includegraphics[scale=0.6]{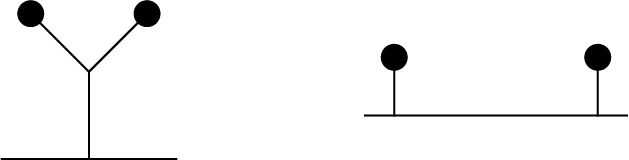}
              \put(-210,18){{\Large $\displaystyle \sum_k$}}              
              \put(-120,18){=}
              \put(-105,18){{\Large $\displaystyle \sum_l$}}
              \put(-164,11){$k$}
              \put(-41,3){$l$}                                    
              \caption{Two-point function crossing symmetry in boundary CFT.}
              \label{csfigure}
            \end{center}
	\end{figure}

The statement of \emph{crossing symmetry} is nothing more than the fact that the two decompositions \eqref{bulkblockdec} and \eqref{boundaryblockdec} should agree, 
\be
\label{crossingsy}
\boxed{
G(\xi) = 1 + \sum_{k} \lambda_k a_{k} \,f_{\text{bulk}}(\Delta_k;\xi) = \xi^\Delta \left( a_{O}^2 + \sum_{l} \mu_l^2 \,f_{\text{bdy}}(\Delta_l;\xi) \right)\, .
}
\ee
A pictorial representation of this equation is shown in figure \ref{csfigure}. 
The aim of this paper is to explore how equation (\ref{crossingsy}) can be used to constrain the space of boundary conformal field theories.

\bigskip

\newpage

\section{The boundary bootstrap in the epsilon expansion}
\label{sec:bootstrapinepsexpansion}
In this section we demonstrate that in a few special cases it is possible to obtain an \emph{analytic} solution of the crossing symmetry equation \eqref{crossingsy}. As we will see below, in this way we can in fact \emph{bootstrap} the outcome of a one-loop computation and recover the order $\eps$ critical exponents of the Wilson-Fisher fixed point! This  is  possible because our solutions turn out to have only one or two blocks in either channel and equation \eqref{crossingsy} reduces to a finite-dimensional linear system. This should be constrasted with the conformal block decomposition for the bulk four-point function, whose asymptotic properties dictate that it always decomposes into an infinite number of conformal blocks \cite{Rattazzi:2008pe}, which makes the problem much harder.
The results in this section therefore highlight  the relative simplicity
of the boundary bootstrap program. At higher orders in the epsilon expansion, the problem becomes infinite-dimensional
even in the boundary case, and more powerful methods will have to be developed.

\subsection{The simplest bootstrap}
\label{subsec:simplestbootstrap}
Let us begin our exploration of the constraining power of the crossing symmetry equation \eqref{crossingsy} by considering the following question: is it possible to satisfy crossing symmetry with just a \emph{single} block in either channel? It turns out that this question can be answered affirmatively and leads to a rederivation of the free-field theory two-point functions.
In formulas, our question becomes whether there exists a solution to the equation
\be
\label{crossingsysingleblock}
1 + \lambda a_\eta \,f_{\text{bulk}}(\eta;\xi) = \xi^\Delta \left( a_{O}^2 + \mu^2 \,f_{\text{bdy}}(\eta';\xi) \right)\,,
\ee
for all $\xi$ and with unknowns $\lambda a_\eta, \eta, \Delta, a^2_{O}, \mu^2$ and $\eta'$. We use $\eta$ and $\eta'$ to denote the dimensions of the single bulk and boundary operator, respectively. 

In order to find a solution we will expand both sides in $\xi$. The bulk conformal blocks \eqref{bulkblock} have a natural series expansion in powers of $\xi$ around $\xi = 0$, which is when we bring the two points close together. On the other hand, the boundary conformal blocks of equation \eqref{boundaryblock} are naturally defined via a series expansion around $\xi = \infty$ where both points approach the boundary.

Now, using standard hypergeometric transformation formulas (see for example \cite{abramowitz+stegun}),
we can expand a boundary block around $\xi = 0$, 
\be
\label{bdyblockat0}
f_{\text{bdy}}(\eta';\xi) = c_1 (1 + \ldots) + c_2 \xi^{1- d/2} (1 + \ldots)\,,
\ee
with the dots representing subleading integer powers of $\xi$ and $c_1$ and $c_2$ certain constants. Substituting this expansion into \eqref{crossingsysingleblock} and simply matching the powers of $\xi$ to those possibly appearing on the left hand side of \eqref{crossingsysingleblock}, we directly find that:
\be
\Delta = \Delta_\phi \equiv \frac{d}{2} - 1\,, \qquad \qquad \eta = 2 \Delta_\phi = d - 2\,.
\ee
This is our first non-trivial result: the scaling dimension $\Delta$ has to be that of a free field $\phi$ and the value of $\eta$ reflects the simple free-field bulk OPE, $\phi \times \phi = \mathbf{1} + \phi^2$.

Our next step is to notice that the bulk block with $\eta = 2 \Delta_\phi$ becomes particularly simple,
\be
f_{\text{bulk}}(2 \D_\phi ;\xi) = \left(\frac{\xi}{\xi + 1} \right)^{\Delta_\phi}\,,
\ee 
and expanding now both sides of \eqref{crossingsysingleblock} around $\xi = \infty$ we find that
\be
1 + \lambda a_\eta \left(1 + \frac{1 - d/2}{\xi} + \ldots \right) = \xi^{\Delta_\phi} \left( a_O^2 + \mu^2 \xi^{- \eta'} \left(1 - \frac{\eta'}{2 \xi} + \ldots \right) \right)\,,
\ee
which allows us to solve for all the other coefficients. We find two possible solutions: 
\be
\begin{split}
+:\qquad \lambda a_\eta = + 1\, , \qquad a_O^2 &= 0\, , \qquad \eta' = \Delta_\phi\, , \qquad \mu^2 = 2\,,\\
-:\qquad  \lambda a_\eta = - 1\, , \qquad  a_O^2 &= 0\, , \qquad \eta' = \Delta_\phi + 1\, , \qquad \mu^2 = \frac{d - 2}{2}\,.
\end{split}
\ee 
Although we have only used the series expansions of the conformal blocks around the endpoints $\xi = 0$ and $\xi = \infty$, it turns out that for the above values of the coefficients the crossing symmetry equation is miraculously satisfied at every order in $\xi$. Therefore, the two functions
\be
\begin{split}
G^+(\xi) &= 1 + f_{\text{bulk}}(2 \Delta_\phi ;\xi) = \xi^{\Delta_{\phi}} \Big(2 f_{\text{bdy}}(\Delta_\phi ;\xi)\Big) = 1 + \left(\frac{\xi}{\xi + 1} \right)^{\Delta_\phi}\, ,
\\
G^-(\xi) &= 1 - f_{\text{bulk}}(2 \Delta_\phi ;\xi) = \xi^{\Delta_{\phi}} \Big( \frac{d-2}{2} f_{\text{bdy}} (\Delta_\phi + 1;\xi) \Big) = 1 - \left(\frac{\xi}{\xi + 1} \right)^{\Delta_\phi}\, ,
\end{split}
\ee
are valid solutions to the crossing symmetry equation \eqref{crossingsy} with just a single block in each channel. Using \eqref{twoptscalar} we find that they correspond to two-point functions of the form:
\be
\vev{\phi(x) \phi(y)} = \frac{1}{(x-y)^{2\Delta_\phi}} \pm \frac{1}{(x-y^r)^{2\Delta_\phi}} \,,
\ee
where $y^r$ is the coordinate vector $y$ reflected in the boundary, so if $y = (\vec y, y^d)$ then $y^r = (\vec y, - y^d)$. This equation informs us that we have derived the two possible two-point functions of a free field on a half-space, with the $+$ sign corresponding to Neumann boundary conditions and the $-$ sign corresponding to Dirichlet boundary conditions.

Let us offer a few more comments on the above solutions. First of all, the bulk-to-boundary OPE is consistent with the boundary conditions. Indeed, the bulk-to-boundary OPE of a free field $\phi$ contains a priori a boundary field $\hat \phi$ and its normal derivative $\del_d \hat \phi$ of dimensions $\Delta_\phi$ and $\Delta_\phi + 1$, respectively. (Notice that these are both $SO(d,1)$ primaries.) As expected, in the Dirichlet case the operator $\hat \phi$ vanishes by the boundary conditions and only the block corresponding to $\del_d \hat \phi$ is present. In the Neumann case the situation is reversed. Finally, the operator $\phi^2$ is the only operator appearing in the bulk channel and the sign of its one-point function is reversed between the two boundary conditions.

\subsection{Order $\epsilon$ bootstrap}

Having obtained the scalar two-point function for the free theory, let us apply the bootstrap technique to the interacting theory in the epsilon expansion. In this section we will allow for $N$ massless scalars with strength $\frac{\l}{4!}(\phi^2)^2$. The $N$-dependence of the free two-point function comes from the overall normalization, so the results of the previous section remain unchanged.
Defining $d=4-\eps$, the Wilson-Fisher fixed point is given by
\be 
\frac{\l_{*}}{16\pi^2}=\frac{3\eps}{N+8} + O(\eps^2)\, .
\ee
We can now write the bootstrap equations as a perturbation series in $\eps$. Following the strategy used in the free case we will assume a finite number of blocks in each channel. In particular, we will consider two non-trivial blocks in the bulk channel and a single block in the boundary channel. This ansatz has some partial justification in Feynman diagrams.
In order for an operator $O$ to appear in the bulk OPE of $\phi$ with itself, the three-point function $\langle \phi \phi  O \rangle$ should be non-zero. For operators of the form $\phi^{2n}$ (ignoring $O(N)$ indices) the only allowed possibilities at order $\eps$ are $\phi^2$ and $\phi^4$. For $n>2$ the correlator is higher order in $\eps$, two or more vertices are needed to contract all the legs. In the boundary channel\footnote{For concreteness we will consider the Neumann case but a parallel analysis can be done for Dirichlet boundary conditions.} we are only considering the operator $\hat{\phi}$, similarly to the bulk case, the bulk-to-boundary OPE between $\phi$ and $\hat{\phi}^{2n+1}$ for $n>0$ is higher order in $\epsilon$.
Let us then proceed to bootstrap the order $\eps$ correlator and comment on the validity of our ansatz at the end of this section.

We want to solve the following equation,
\be 
\label{epsexponeloop}
1 + \l a_{\phi^2} f_{\text{bulk}}(\Delta_{\phi^2};\xi) +  \l a_{\phi^4}  f_{\text{bulk}}(\Delta_{\phi^4};\xi) = \mu^2\xi^{\Delta_{\phi}}f_{\text{bdy}}(\Delta_{\hat{\phi}},\xi)\, .
\ee
Because we are working perturbatively we will write all coefficients as a power series in $\eps$. For the spacetime dimension $d$ and the external dimension conformal
dimension $\Delta_\phi$ we have
 \be 
 \begin{split}
 d & =  4-\eps\, ,
 \\
 \Delta_\phi & =  \frac{d}{2}-1 + \delta \Delta_\phi \eps + O(\eps^2)\, .
 \end{split}
 \ee
 For the internal conformal dimensions we write,
  \be 
 \begin{split}
 \Delta_{\phi^2} & =  d-2 + \delta \Delta_{\phi^2} \eps + O(\eps^2) \, ,
  \\
 \Delta_{\phi^4} & =  2d-4 + \delta \Delta_{\phi^4} \eps + O(\eps^2) \, ,
 \\
 \Delta_{\hat{\phi}} & =  \frac{d}{2}-1 + \delta  \Delta_{\hat{\phi}} \eps + O(\eps^2) \, .
 \end{split}
 \ee
 Finally, for the coefficients multiplying the blocks,
  \be 
 \begin{split}
 \l a_{\phi^2} & =  1 +  \delta \l a_{\phi^2}\eps + O(\eps^2) \, ,
 \\
 \l a_{\phi^4} & =  \delta \l a_{\phi^4}\eps + O(\eps^2) \, ,
 \\
 \mu^2 & =  2 +  \delta \mu^2 \eps + O(\eps^2) \, ,
 \end{split}
 \ee
where the quantities denoted by ``$\delta$'' correspond to deviations from the free-field solution. For example, $\l a_{\phi^4}$ has only a correction term since it is not present in the free theory.
We will again use the transformation formulas that led to (\ref{bdyblockat0}) in order to expand the boundary blocks around $\xi=0$. The procedure now is the same as before, we Taylor expand both sides of the equation and match equal powers of the parameter $\xi$. As in the free case, after matching the first few coefficients, equation (\ref{epsexponeloop}) is solved to all orders in $\xi$.
The order $\epsilon$ solution is,
\be
\begin{aligned}
\delta \Delta_\phi & = 0\, , \qquad \qquad &  \delta \Delta_{\phi^2}  & = 2\a\, , \qquad \qquad &   \delta  \Delta_{\hat{\phi}}   & = -\a\, , 
\\
 \delta \l a_{\phi^2} & = \a\, , \qquad \qquad & \delta \l a_{\phi^4}  & = \frac{\a}{2}\, , \qquad \qquad & \delta \mu^2  & = 0\, , & 
\end{aligned}
\ee
where $\a$ is an arbitrary coefficient. The zero one-loop anomalous dimension for $\phi$ is not a surprise, the anomalous dimension of $\phi^2$ is also well known and can be used to fix the value of $\a$,
\be 
\a = \frac{1}{2}\left(\frac{N+2}{N+8} \right)\, .
\ee
The first order corrections to the OPE coefficients of the $\phi^2$ and $\phi^4$ blocks are positive, while the order $\eps$ correction to $\mu^2$ is zero, as expected from Feynman diagrams. We find a negative anomalous dimension for the boundary operator corresponding to $\hat \phi$. 
The anomalous dimension for $\phi^4$ does not enter the equations at this order in the expansion.
The complete corrected two-point function is then
\begin{align}
G^+_{\phi \phi} &= 1 + \left( \frac{\xi}{\xi + 1}\right)^{1 - \frac{\eps}{2}} + \frac{\eps}{2} \left(\frac{N+2}{N+8} \right) \Big( \frac{\xi}{\xi + 1} \log(\xi) + \log(\xi + 1) \Big) + O(\eps^2) \nonumber \\
&= 1 + \Big(1 + \frac{\eps}{2} \left(\frac{N+2}{N+8} \right) \Big)f_{\text{bulk}}(2-\eps+ \eps \left(\frac{N+2}{N+8} \right) ;\xi) + \frac{\eps}{4} \left(\frac{N+2}{N+8} \right) f_{\text{bulk}} (4 ; \xi) + O(\eps^2) \nonumber \\
&= \xi^{1 - \frac{\eps}{2}} \Big( 2 f_\text{bdy} (1 - \frac{\eps}{2} - \frac{\eps}{2} \left(\frac{N+2}{N+8} \right); \xi) \Big) + O(\eps^2)\, ,
\end{align}
where the $+$ sign indicates Neumann boundary conditions. An analogous calculation can be done for the Dirichlet case. We simply quote the result:
\begin{align}
G^-_{\phi \phi} &= 1 - \left( \frac{\xi}{\xi + 1}\right)^{1 - \frac{\eps}{2}} + \hf \eps \left(\frac{N+2}{N+8} \right) \Big( - \frac{\xi}{\xi + 1} \log(\xi) + \log(\xi + 1) \Big) + O(\eps^2) \nonumber \\
&= 1 - \Big(1 - \hf \eps \left(\frac{N+2}{N+8} \right)  \Big)f_{\text{bulk}}(2 -\eps + \eps \left(\frac{N+2}{N+8} \right)  ;\xi) + \frac{\eps}{4} \left(\frac{N+2}{N+8} \right) f_{\text{bulk}} (4 ; \xi) + O(\eps^2) \nonumber \\
&= \xi^{1 - \frac{\eps}{2}} \Big( \Big(1 - \frac{\eps}{2} + \frac{\eps}{2} \left(\frac{N+2}{N+8} \right) \Big) f_\text{bdy} (2-\frac{\eps}{2} -\frac{\eps}{2} \left(\frac{N+2}{N+8} \right) ; \xi) \Big) + O(\eps^2)\, ,
\end{align}
which features only minor changes with respect to the previous case. Comparison of these expressions with the explicit calculation of \cite{McAvity:1995zd} shows perfect agreement. We have used the bootstrap equations to obtain a one-loop result!

Let us now return to our original ansatz.
We did not consider primary operators with derivatives acting on the $\phi$, which we denote schematically by $\Box^k \phi^2$ and $\Box^k \phi^4$. For the first family, we can never have $\pd_\mu \pd_\mu$ acting on the same field, because the equations of motion imply $\pd_\mu \pd_\mu \phi \sim \eps \phi^3$ and the operator is not really of the form $\Box^k \phi^2$. The only possibility is to have $\pd_{\mu_1}\pd_{\mu_2} \ldots \pd_{\mu_k} \phi \pd_{\mu_1}\pd_{\mu_2} \ldots \pd_{\mu_k} \phi$, but these operators are conformal descendants, and their contribution is already taken into account by the $\phi^2$ block. For the second family, the equations of motion argument still holds, but not all operators are conformal descendants. In fact, there is an infinite number of primaries of the schematic form $\Box^k \phi^4$.\footnote{This statement can be checked using conformal characters.} Our original ansatz was thus incomplete, we should have added an infinite number of blocks to the left-hand side of equation (\ref{epsexponeloop}) with tree level dimension $\Delta_k = 2(d-2)+2k$. As we obtained the correct answer, it is clear that these operators do not appear at one loop. We believe that this is due to the vanishing of the three-point functions $\langle \phi \phi \Box^k \phi^4\rangle$ for $k>0$,
a fact which should follow from the higher-spin Ward identities of the free theory.

Starting at order $\eps^2$, crossing symmetry can no longer be solved with a finite number of blocks.  It would be
nice to find more powerful analytic techniques to deal with the infinite-dimensional linear system, and
develop a bootstrap apprach to the all-order epsilon expansion.
 At each order
a new infinite family of bulk primary operators appears. Perhaps the constraints of sligthly broken higher-spin symmetry \cite{Maldacena:2011jn, Maldacena:2012sf}
could help
in organizing the information contained in (\ref{crossingsy}). We leave this as an intriguing direction for future work,
and devote the rest of the paper to  numerical investigations. 

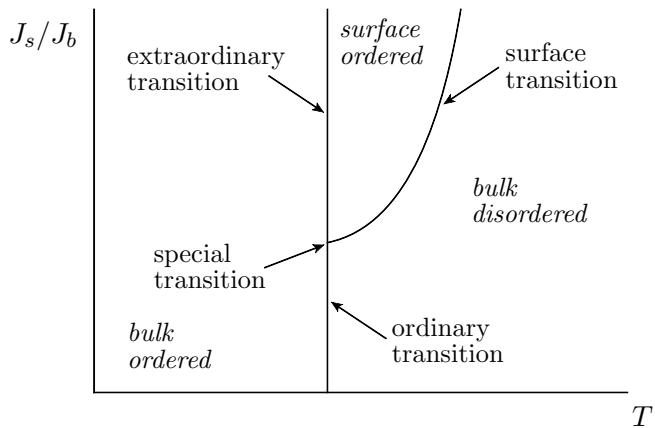
\begin{figure}[H]
             \begin{center}           
              \includegraphics[scale=0.9]{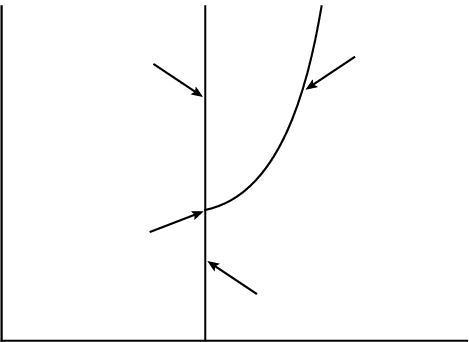}
              \put(-190,125){{\small extraordinary}}   
              \put(-190,115){{\small transition}}
               \put(-180,50){{\small special}}   
              \put(-180,40){{\small transition}}   
               \put(-90,22){{\small ordinary}}   
              \put(-90,12){{\small transition}} 
               \put(-47,126){{\small surface}}   
              \put(-47,116){{\small transition}}     
              \put(-110,135){{\small \textit{surface}}}   
              \put(-110,125){{\small \textit{ordered}}}   
              \put(-60,75){{\small \textit{bulk}}}   
              \put(-60,65){{\small \textit{disordered}}}     
              \put(-190,20){{\small \textit{bulk}}}   
              \put(-190,10){{\small \textit{ordered}}} 
              \put(1,-12){$T$} 
              \put(-235,133){$J_s / J_b$}                         
              \caption{Phase diagram for the surface critical behavior of the Ising model in dimension $2 < d <4$. Temperature is plotted on the horizontal axis and the (relative) surface interaction strength on the vertical axis. The extraordinary transition disappears for $d=4$, while the special transition is absent in $d=2$.
              }
              \label{phasediagram}
            \end{center}
\end{figure}

\paragraph{Statistical mechanics intermezzo\\}
In the study of critical systems with a boundary it is well-known that Neumann boundary conditions for the Landau-Ginzburg  field $\phi$
(which corresponds to the bulk spin operator $\sigma$)
describe  the so-called \emph{special transition}, while Dirichlet boundary conditions describe the \emph{ordinary transition}. The phase diagram of the Ising model in the presence of a boundary is shown in figure \ref{phasediagram}.

In our investigations the bulk is always critical so we are always on the vertical line in figure \ref{phasediagram}.  For weak boundary interactions one finds there the \emph{ordinary} transition where the boundary simply orders at the same temperature as the bulk. In the presence of strong boundary interactions the boundary can however order at a higher temperature than the bulk. The bulk transition where the boundary is already ordered is then called an \emph{extraordinary} transition. In this case the $\mathbb{Z}_2$ symmetry of the Ising model is broken, as 
 $\phi$ should acquires a one-point function of the form \eqref{oneptscalar}. The extraordinary transition cannot be described in free-field theory
 (such a one-point function does not satisfy the free equations of motion), but it appears at first order in the Wilson-Fisher fixed point in $4-\epsilon$ dimensions, see appendix \ref{subsec:phiextraord}.
Finally, there is a critical boundary interaction strength where the boundary and bulk critical temperature just coincide which is the \emph{special} transition. We refer the reader to  \cite{diehl, cardy} for  introductions to boundary critical phenomena.

The BCFT associated to the extraordinary transition is the most ``stable'' as there are no relevant boundary scalar operators.
In fact it is believed that its lowest-dimensional boundary scalar is the ``displacement operator'' $\hat T_{dd}$, which is the boundary
limit of the bulk stress tensor with both indices pointing in the direction normal to the boundary. The displacement operator
has protected conformal dimension exactly equal to $d$, and it is thus irrelevant on the $(d-1)$-dimensional boundary.
The BCFTs associated to the ordinary and special transitions preserve the $\mathbb{Z}_2$ symmetry, which thus remains a good
quantum number for boundary operators.
 The boundary spectrum of the BCFT associated to
the ordinary transition contains  a single  relevant  scalar operator which is $\mathbb{Z}_2$ odd,
and corresponds to $\partial_d \hat \phi$ in the Landau-Ginzburg description. Finally there are {\it two} relevant scalars
in the BCFT  for the special transition, one $\mathbb{Z}_2$ odd and the other $\mathbb{Z}_2$ even, corresponding 
respectively to $\hat \phi$ and $\hat \phi^2$.  

In $d=2$, the extraordinary transition is associated to the Cardy boundary states $| \mathbf 1 \rangle \! \rangle$ and $| \veps \rangle \! \rangle$ labelled by the identity and the energy, respectively. We have
\be
\label{cardystatesextr}
\begin{split}
| \mathbf 1 \rangle\! \rangle &=  \frac{1}{\sqrt 2} |  \mathbf 1  \rangle   + \frac{1}{\sqrt{2}}  | \veps  \rangle   + \frac{1}{\sqrt[4]{2}} | \sigma \rangle\, ,\\
| \veps \rangle\! \rangle &=  \frac{1}{\sqrt 2} | \mathbf 1  \rangle   + \frac{1}{\sqrt{2}}  | \veps  \rangle   -  \frac{1}{\sqrt[4]{2}} | \sigma \rangle\, ,
\end{split}
\ee
where the kets on the right-hand side denote Ishibashi states. We see that the two states are physically equivalent since they are being related by $\mathbb{Z}_2$ conjugation. The ordinary transition is associated instead to the Cardy boundary state $| \sigma \rangle\!\rangle$ labelled by the spin,  which is given by
\be
\label{cardystatesord}
| {\sigma} \rangle\! \rangle =  | \mathbf 1 \rangle - | \veps \rangle\,.
\ee
There is no $2d$ BCFT associated to the special transition, since the one-dimensional boundary cannot order dynamically at non-zero temperature and so the surface transition is absent.

\section{Numerical results for scalars}
\label{sec:scalarnumerics}
Despite the promising results obtained at zeroth and first order  in the $\eps$ expansion, currently no good analytic tools are available for the exploration of the general space of solutions of the crossing symmetry equation  \eqref{crossingsy}.
Therefore we have to resort to numerical approaches. In this section we adapt the numerical methods of \cite{Rattazzi:2008pe} to our case and derive
exclusion curves for operator dimensions and OPE coefficients.

The results we obtain below will depend sensitively on some assumptions about the boundary operator spectrum and thereby fall naturally into different categories related to the different possible boundary conditions. Following \cite{ElShowk:2012ht} we will focus mainly on correlation functions of the $\sigma$ operator in the three-dimensional Ising model, whose possible boundary conditions were presented in figure \ref{phasediagram}. For reasons to be discussed in subsection \ref{subsec:scalarnumerimpl}, our focus will be on the \emph{special} and \emph{extraordinary} transitions, which will respectively be discussed in subsections \ref{subsec:special} and \ref{subsec:extraord} below. The relevant bulk and boundary operator product expansions and scaling dimensions are summarized in table \ref{tab:critexpising}. For $d=4$ there are several operators that do not appear in OPE and we indicated this with a dash. The quoted values for the Ising model in $d=3$ are of course approximate, but good enough for the numerical precision of this paper.  We were unable to find a reliable estimate of the dimension of the $\hat \sigma'$ operator for the special transition.

\begin{table}[h!]
\centering
\hfill
\begin{tabular}[t]{|c||c|c|c|}
\hline
\multicolumn{4}{|c|}{bulk}\\
\hline \hline
\multicolumn{4}{|c|}{$\sigma \times \sigma  = 1 + \veps + \veps' + \veps'' + \ldots$}\\
\hline \hline
$d$ & $\,\,\,\,\,\,$2$\,\,\,\,\,\,$ & 3 & $4$ \\
\hline \hline
$\D_\sigma$ & $\frac{1}{8}$ & 0.5182(3) & 1\\
\hline
$\D_\eps$ & 1 & 1.413(1) & 2\\
\hline
$\D_{\eps'}$ & 4 & 3.84(4) & -\\
\hline
$\D_{\eps''}$ & 8 & 4.67(11) & -\\
\hline
\end{tabular} \hfill
\begin{tabular}[t]{|c||c|c|}
\hline
\multicolumn{3}{|c|}{special}\\
\hline \hline
\multicolumn{3}{|c|}{$\sigma  = \hat \sigma + \hat \sigma' + \ldots$}\\
\hline \hline
$d$ &  3 & $\,\,\,\,\,$4$\,\,\,\,\,$ \\
\hline \hline
$\D_{\hat \sigma}$ & 0.42 & 1\\
\hline
$\D_{\hat \sigma'}$ &  ? & -\\
\hline
\end{tabular}\hfill
\begin{tabular}[t]{|c|}
\hline
extraordinary\\
\hline \hline
$\sigma  = 1 + \hat T_{dd} + \ldots$\\
\hline
\end{tabular}
\hfill
\caption{\label{tab:critexpising}Bulk and boundary operator product expansions and operator dimensions in the Ising model in various dimensions. There is no special transition in two dimensions. For the extraordinary transition the first boundary operator is $\hat T_{dd}$ whose dimension is always equal to the spacetime dimension $d$. The results for $d=3$ are approximate and were obtained from \cite{ElShowk:2012ht,Diehlpaper} whereas the results for $d=2$ and $d=4$ can be found in the appendices.}
\end{table}

\subsection{Implementation}
\label{subsec:scalarnumerimpl}
Let us review how to implement the optimization problem numerically. The following techniques were explained in great detail in \cite{Rattazzi:2008pe,Poland:2010wg} so we shall be brief. 
We start by isolating the contribution of the identity operator in equation \eqref{crossingsy},
\be
1  =  -\sum_{k} \lambda_k a_{k} \,f_{\text{bulk}}(\Delta_k;\xi) + \xi^{\Delta_{\text{ext}}} \left( a_O^2 + \sum_{l} \mu_l^2 \,f_{\text{bdy}}(\Delta_l;\xi) \right)\, ,
\ee
and introduce the compact notation,
\be
\label{sumrule}
1 = \sum_{\Delta} p_{\Delta}F_{\Delta}(\xi)\, ,
\ee
where
\bea 
p_{\Delta} & = & \left(\lambda_k a_{k}\,, a_O^2\,,\mu_l^2\right)\, ,
\\
F_{\Delta}(\xi) & = & \left(-f_{\text{bulk}}(\Delta_k;\xi)\, , \xi^{\Delta_{\text{ext}}} \, ,\xi^{\Delta_{\text{ext}}} f_{\text{bdy}}(\Delta_l;\xi)\right)\, .
\label{generalF}
\eea
With these definitions equation (\ref{sumrule}) is analogous to the sum rule of \cite{Rattazzi:2008pe}. There is however a crucial difference between the boundary problem that we are studying compared to the four-point function crossing symmetry of \cite{Rattazzi:2008pe}: even assuming unitarity (as we shall always do) 
 the coefficients $p_{\Delta}$ are not all guaranteed to be positive.
They are certainly positive in the boundary channel, since they are squares of real numbers, but in the bulk channel the combination
$\lambda_k a_k$ is not manifestly positive. Indeed it is not difficult to find counterexamples (such as a free scalar with Dirichlet boundary conditions).
 In the following, we will \textit{assume} positivity for the bulk expansion such that $p_{\Delta} \ge 0$ as in the four-point function case. The conjecture  is that 
 for a given bulk CFT, there exists a choice of boundary conditions 
  that exhibits positivity. In the Ising model, the ordinary transition is excluded from our analysis, since both signs occur in the bulk expansion (as can be demonstrated
  in $d=2$ and in $d = 4 - \epsilon$ dimensions). 
   We will however assume positivity for the special and the extraordinary transitions. This assumption is supported by the results in the previous section as well as in the appendices. We have found positivity of the bulk block coefficients around $d=4$, both for the free field and the Wilson-Fisher fixed point at order $\epsilon$, as well as in $d=2$ where it is a consequence of the positivity of the first two coefficients in the first line of \eqref{cardystatesextr}. In appendix \ref{sec:ONlargeN} we also found that the coefficients for the special transition are positive in the $O(N)$ model at large $N$ for any dimension.

We are now ready to start extracting information from the sum rule (\ref{sumrule}). The simplest possible bound can be obtained as follows:
We allow for the bulk spectrum to span all possible values consistent with unitarity,
\be 
\label{bulkgap}
\Delta_{\text{bulk}} \ge \frac{d-1}{2}\, ,
\ee
while restricting the boundary spectrum to be greater than a given value,
\be 
\label{bdygap}
\Delta_{\text{bdy}} \ge \Delta_{\text{min}}\, .
\ee
Then, we consider a functional $\Lambda$ with the following properties,
\bea 
\Lambda(1)& < & 0\, ,
\\
\Lambda(F_\D)& \ge & 0\, ,
\eea
where, according to our definitions, $F_\D$ stands for any of the blocks appearing in \eqref{generalF} with scaling dimensions obeying \eqref{bulkgap} and \eqref{bdygap}. If such a functional is found, equation (\ref{sumrule}) becomes inconsistent and we can rule out that particular CFT. The idea then is to see how low we can push $\Delta_{\text{min}}$.

Before implementing the machinery of linear functionals we need to choose a set of ``coordinates'' in our function space. We will parametrize the blocks by an infinite vector of derivatives  $\{F_\D^k\}$ evaluated at $\xi=1$,
\be 
F^k_\D =  \left. \frac{\partial^k F_\D(\xi)}{\partial \xi^k}\right|_{\xi=1}\,,
\ee
and crossing symmetry becomes now an infinite set of algebraic equations. In order to make the problem numerically tractable we will discretize the spectrum of bulk and boundary dimensions and consider a maximum number of derivatives. With this truncation we have an optimization problem with a finite dimensional set of inequalities, this is an example of a \textit{linear program}. In order to solve the linear programs we used the \texttt{Mathematica} routine \texttt{LinearProgamming} and the IBM ILOG CPLEX Optimizer. In all our plots below we used a grid of $\delta_k=0.01$ and a total of 15 derivatives.

\subsection{Special transition}
 \label{subsec:special}
In the following we present our numerical results for the special transition. 
The one-point function of the bulk spin operator $\sigma$ vanishes since the $\mathbb Z_2$ symmetry is unbroken by the (Neumann) boundary conditions. As we have emphasized in the previous subsection, positivity of the bulk channel coefficients will be a working assumption.

\subsubsection{Simplest bound for the boundary channel}
\begin{figure}[H]
             \begin{center}     
              \includegraphics[scale=0.6]{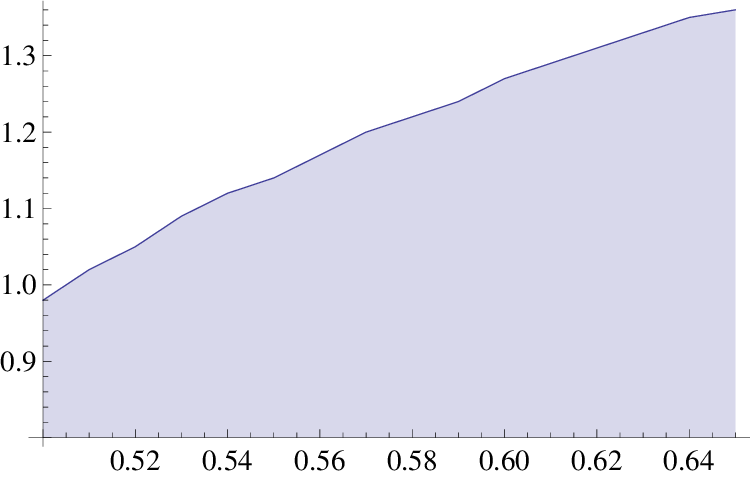}     
               \put(2,-5){$\Delta_{\text{ext}}$}    
               \put(-245,128){$\Delta_{\text{bdy}}$}                   
              \caption{Upper bound for the first boundary operator in the special transition.}
              \label{NoGap}
            \end{center}
	\end{figure}

Let us start by plotting the simplest possible bound of the form described above. Our only assumption for the bulk spectrum will be the three-dimensional unitarity bound, $\Delta_{\text{bulk}} \ge 0.5$, but otherwise bulk operators of any dimension are allowed to appear in the OPE. Crossing symmetry and positivity however imply that the conformal dimension of the lowest dimension boundary operator cannot be arbitrary. Instead, we found that depending on the external dimension the first boundary operator has to lie below the curve of figure \ref{NoGap}.

Although this is a correct 
bound, we should mention the following caveat: The bulk block 
 blows up at the unitarity bound and our more precise assumption for the bulk spectrum was actually 
$\D_\text{bulk} \ge 0.5 + 10^{-6}$. Unfortunately, it turns out that the numerics are quite sensitive around this point. For example, the bound becomes much stronger if we change our assumptions on the bulk spectrum to $\Delta_\text{bulk} \ge 0.51$.
Because of this, we do not consider this plot to be physically very relevant but it serves as a good warm-up example before tackling the most interesting cases below. 

\subsubsection{Improved bound for the boundary channel}

\begin{figure}[H]
             \begin{center}                
              \includegraphics[scale=0.6]{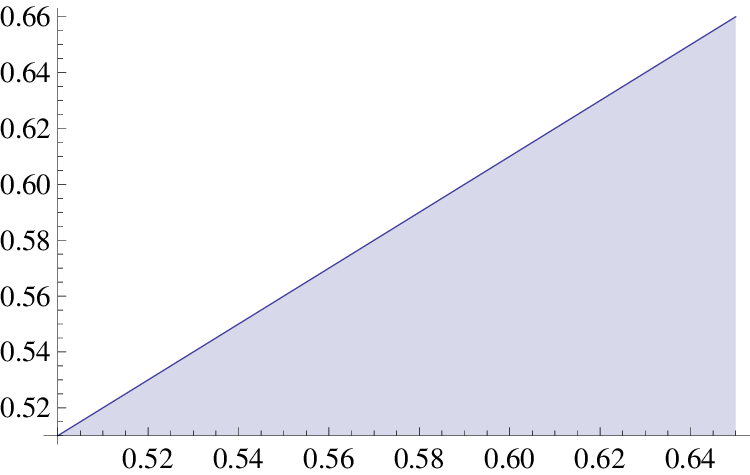}    
              \put(2,-5){$\Delta_{\text{ext}}$}    
               \put(-245,128){$\Delta_{\text{bdy}}$}                         
              \caption{Improved bound for the first boundary operator in the special transition. The bulk spectrum is assumed to satisfy $\Delta_\text{bulk} \ge 2 \Delta_{\text{ext}}$.}
              \label{Gap2eta}
            \end{center}
\end{figure}

The boundary bound obtained above can be improved by making further assumptions. In the bulk channel decomposition of a scalar two-point function we expect, on physical grounds, a ``gap'' between the unitarity bound and the conformal dimension of the first operator appearing in the bulk OPE. For example, according to table \ref{tab:critexpising}, in the three-dimensional Ising model the first bulk operator appearing in the OPE of the spin operator $\s$ is the energy operator $\veps$ with $\Delta_{\veps} = 1.41$, far above the unitarity bound. Clearly, allowing for the bulk spectrum to go all the way down to the unitarity bound is very unphysical. In figure \ref{Gap2eta} we present an improved bound in which we assumed that the bulk spectrum satisfies $\Delta_\text{bulk} \ge 2 \Delta_{\text{ext}}$.

Our solution seems to indicate that the bound cannot go below the straight line where $\Delta_{\text{bdy}} = \Delta_{\text{ext}}$. The reason for this is the trivial solution $(x_1-x_2)^{-2\Delta_{\text{ext}}}$ which we discuss in appendix \ref{subsec:trivialsol}. This two-point function contains no non-trivial bulk blocks and thus effectively has an infinite gap in the bulk spectrum. On the other hand, it also has a boundary channel expansion which starts with a block of dimension $\Delta_{\text{ext}}$ and our bound of course cannot get past this particular solution. In a sense, the bound is optimal in this case, going down until it hits a known solution to crossing symmetry.

For the Ising model the dimension of the first boundary operator has a value of $\sim 0.42$ and is well inside the allowed region of figure \ref{Gap2eta}. Ideally, we would have found a plot with some striking feature around this value, like the kink of \cite{ElShowk:2012ht}. However, in our case the trivial solution is standing in the way. A qualitative explanation for this difference appears in the epsilon expansion results. Namely, the anomalous dimension of the $\veps$ operator (which is $\phi^2$ in $d=4$) is \emph{positive} at one loop, so the Ising model lies \emph{above} any trivial (mean field-like) solutions for the bulk four-point function. On the other hand, the one-loop anomalous dimension of the first boundary operator is \emph{negative}, so we end up \emph{below} the trivial solution. This was of course largely a coincidence - we are not aware of any fundamental reason requiring these anomalous dimensions to have a definite sign. Some effort was made in order to circumvent the trivial solution but we did not succeed in obtaining reliable ``kinks'' that highlight the presence of the Ising model. 

We would like to stress however that our plot is still teaching us something very non-trivial: the lowest boundary dimension can never be greater than the external dimension. Interestingly, this result precisely implies that the bulk-to-boundary OPE is never regular, see equation \eqref{crossingsy}. It would be very interesting to find a more direct argument for this result ---perhaps even one that does not rely on our specific assumptions.

\subsubsection{Bounding the second boundary operator in the Ising model}	

\begin{figure}[H]
             \begin{center}        
         
              \includegraphics[scale=0.6]{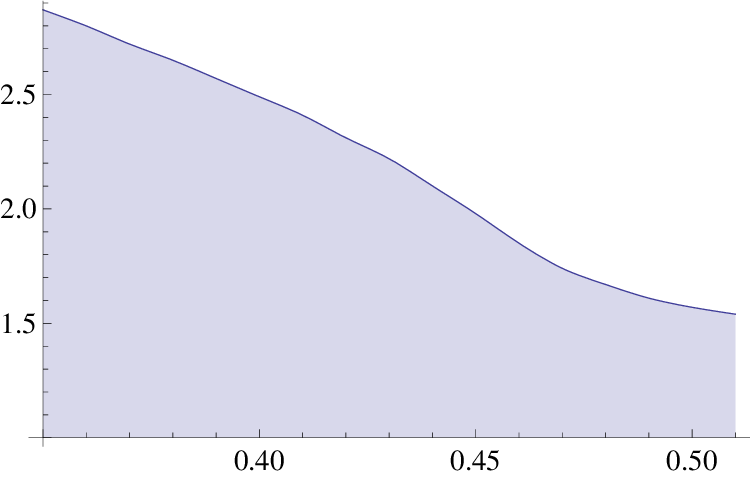}
              \put(2,-5){$\Delta^{(1)}_{\text{bdy}}$}    
              \put(-245,128){$\Delta^{(2)}_{\text{bdy}}$}                
              \caption{Upper bound for the dimension of the second boundary operator in $\langle \sigma \sigma \rangle$ as a function of the dimension of the first boundary operator.}
              \label{sndbdyoperatorss}
            \end{center}
	\end{figure}

Our assumptions in the previous section were almost minimal, and the result is a general bound valid on the space of BCFTs. In this section we will take a closer look at the three-dimensional Ising model and attempt to bound the second boundary operator. We will do so for both the $\langle \s \s \rangle$ and $\langle \veps \veps \rangle$ correlators. Using the results from table \ref{tab:critexpising} we can assume that
\be
\begin{split}
\Delta_{\text{ext}} &= 0.518\, ,\\
\Delta_{\text{bulk}} & \ge 1.41\, ,\\
\Delta^{(1)}_{\text{bdy}} & \sim 0.42\, ,\\
\Delta^{(2)}_{\text{bdy}} & \ge \Delta^{(2)}_{\text{min}}\, .\\
\end{split}
\ee 
In the boundary channel the first block corresponds to $\hat \sigma$. We assume that it sits isolated at $\Delta^{(1)}_{\text{bdy}} \sim 0.42$ and that all the subsequent blocks have a scaling dimension greater than $\Delta^{(2)}_{\text{min}}$. Proceeding as before we push $\Delta^{(2)}_{\text{min}}$ as high as possible until the CFT becomes inconsistent. This will give us an upper bound for the dimension of the second operator $\hat \sigma'$, only valid for the $\langle \s \s \rangle$ correlator of the 3d Ising model. Because $\sim 0.42$ is our less precise value we will explore a range around this number. Our result is shown in figure \ref{sndbdyoperatorss}.
	
\begin{figure}[H]
             \begin{center}        
         
              \includegraphics[scale=0.6]{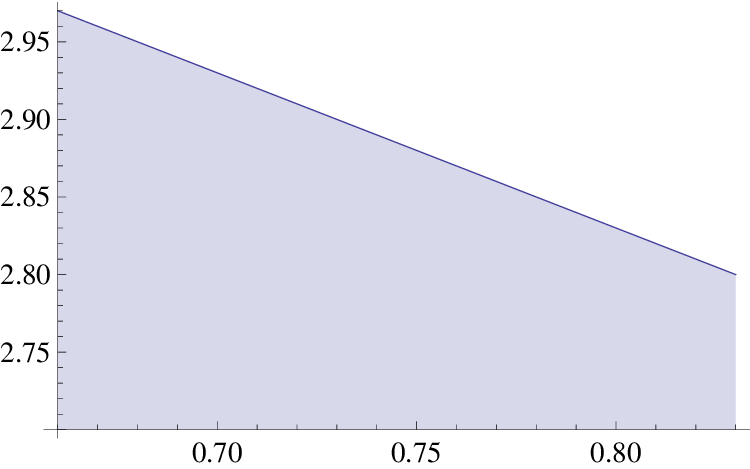}
              \put(2,-5){$\Delta^{(1)}_{\text{bdy}}$}    
              \put(-245,128){$\Delta^{(2)}_{\text{bdy}}$}                 
              \caption{Upper bound for the second boundary operator in  $\langle \varepsilon \varepsilon \rangle$ as a 
              function of the first boundary operator.}
              \label{sndbdyoperatoree}
            \end{center}
	\end{figure}

The same can be done for the $\langle \varepsilon \varepsilon \rangle$ correlator. The statistical mechanics data \cite{Diehlpaper} in this case are
\be
\begin{split}
\Delta_{\text{ext}} &= 1.41\, ,\\
\Delta_{\text{bulk}} & \ge 3.80\, ,\\
\Delta^{(1)}_{\text{bdy}} & \sim 0.75\, ,\\
\Delta^{(2)}_{\text{bdy}} & \ge \Delta^{(2)}_{\text{min}}\, .\\
\end{split}
\ee 
and the resulting bound is shown in figure \ref{sndbdyoperatoree}.

Unfortunately, we were unable to find reliable estimates of the scaling dimensions of the second boundary operators in the statistical mechanics literature. It would of course be interesting to compare our values with \emph{e.g.} a two-loop computation for the Wilson-Fisher fixed point.

\subsection{Extraordinary transition}
\label{subsec:extraord}
In the extraordinary transition the boundary identity operator is always present, so bounding the lowest boundary dimension
is not an interesting exercise in this case. The second boundary scalar operator is expected  to be $\hat T_{dd}$,  the energy momentum tensor with indices in the normal direction,
evaluated on the boundary. This operator is always present in the boundary spectrum and has conformal dimension exactly equal to $d$, see \cite{McAvity:1993ue} for details. Having so much information about the boundary channel we would like to address the following question: can we bound the bulk spectrum using the boundary bootstrap? We will show below that this is indeed possible, although our bound is weaker than the one obtained in \cite{ElShowk:2012ht} who used the crossing symmetry equations for the bulk four-point function.

\subsubsection{Bound for the bulk channel}

\begin{figure}[H]
             \begin{center}               
              \includegraphics[scale=0.6]{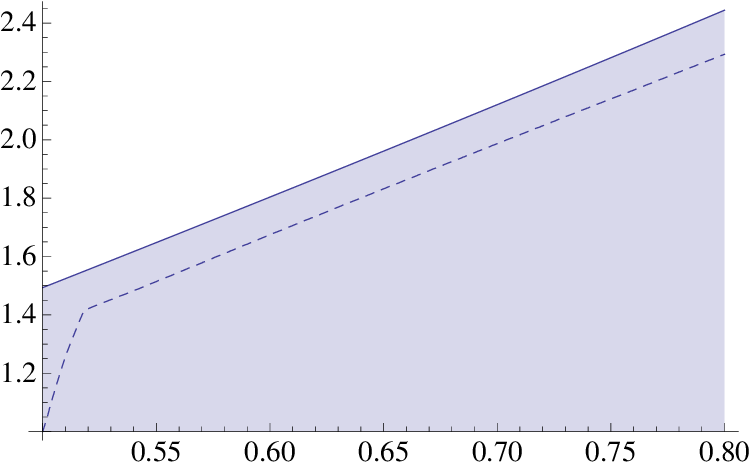}
              \put(2,-5){$\Delta_{\text{ext}}$}    
              \put(-245,128){$\Delta_{\text{bulk}}$}                             
              \caption{Bulk bound for the extraordinary transition as a function of the external dimension. The dashed line corresponds to the (stronger) bound obtained in \cite{ElShowk:2012ht} using the bulk crossing symmetry equations.}
              \label{bulkextra}
            \end{center}
\end{figure}

The assumptions for the extraordinary transition are
\be
\begin{split}
\Delta_{\text{bulk}} & \ge \Delta_{\text{min}}\, .\\
\Delta^{(1)}_{\text{bdy}} &= 0\, ,\\
\Delta^{(2)}_{\text{bdy}} & \ge d\, ,\\
\end{split}
\ee
where we used a notation familiar from the previous subsection.  The fact that $\Delta^{(1)}_{\text{bdy}}  = 0$ corresponds to the boundary identity operator which sits isolated, and we then allow for any operator with a dimension greater than (or equal to) $d$ to be present in the boundary channel. $\Delta_{\text{min}}$ is the lowest bulk dimension and the quantity we want to bound. In figure \ref{bulkextra} we plot our bound as a function of the external dimension.     
Because figure \ref{bulkextra} can be directly compared with the bound of \cite{ElShowk:2012ht} we have superimposed their result on our plot. We can see that the bound obtained using the boundary bootstrap is qualitatively different, it is weaker and has no kink at the Ising point. Since we successfully found an ``optimal'' bound for the boundary spectrum in the previous subsection, it is surprising that our bulk bound does not exhibit any of the expected features.

There are two possible explanations for the discrepancy seen in figure \ref{bulkextra}. First, there may be a spurious solution to crossing symmetry that we have not found yet and that prevents the bound from going lower. If such a solution exists then it would be interesting to understand whether it corresponds to a full-fledged BCFT or not. Notice that this solution would appear to violate the bound of \cite{ElShowk:2012ht} but this may be due to the fact that certain operators do not get one-point functions and therefore do not appear in our bulk block expansion. The second explanation is that our numerics are not precise enough and that we would be able to lower the bound by increasing our numerical precision. We offer some comments on this second possibility below.

\subsubsection*{Bulk bound for arbitrary $d$}
\begin{figure}[H]
             \begin{center}               
              \includegraphics[scale=0.6]{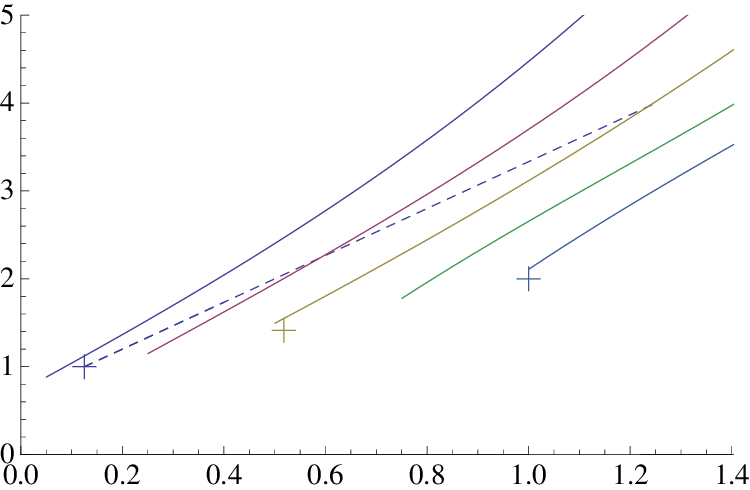}
               \put(-51,142){{\footnotesize$d=2$}}
               \put(-15,142){{\footnotesize$d=2.5$}}
               \put(2,127){{\footnotesize$d=3$}}
               \put(2,111){{\footnotesize$d=3.5$}}
               \put(2,97){{\footnotesize$d=4$}}
               \put(2,-5){$\Delta_{\text{ext}}$}    
               \put(-245,128){$\Delta_{\text{bulk}}$}                             
              \caption{Bulk bound for different spacetime dimensions in the extraordinary transition. We highlighted the Ising model in various dimensions with the crosses. The dashed line is a specific solution for $d=2$ which interpolates through the minimal models, see  appendix \ref{sec:minimalmods}.}
              \label{bulkboundarbitraryd}
            \end{center}
\end{figure}
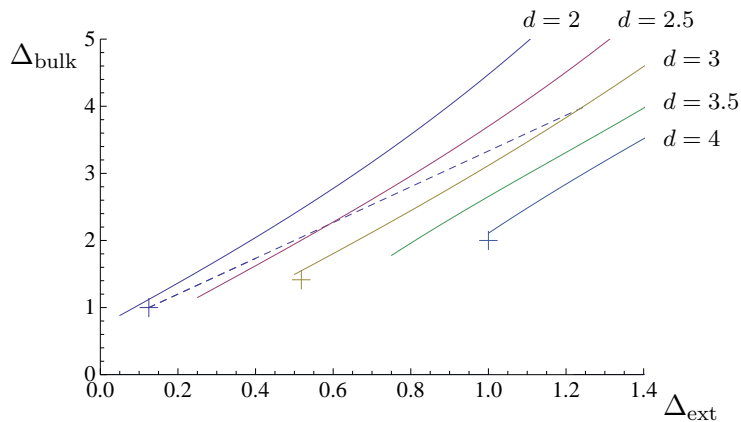

One of the advantages of studying the boundary problem is that the blocks are an analytic function of $d$. In figure \ref{bulkboundarbitraryd} we plot the bulk bound obtained above for different dimensions including non-integer values.

The bound we find is always significantly different from any known solutions to crossing symmetry. In particular, in the figure we have shown the line interpolating through the minimal models in $d=2$ and the Ising model for the integral dimensions. Again, it would be interesting to understand if this is due to our finite numerical precision or whether there exist `spurious' solutions to the crossing symmetry equations at the current bounds. 

\subsubsection{Upper bound for $\hat T_{dd}$ OPE coefficient}
	
\label{subsubsec:boundopecoeffscalar}
The method of linear functionals can also be used to bound OPE coefficients. In \cite{Caracciolo:2009bx} a universal upper bound for the OPE coefficient of three scalars was found using the four-point function bootstrap. The same technique was used in \cite{Rattazzi:2010gj,Poland:2010wg} to obtain an upper bound for the OPE coefficient of the stress tensor. This coefficient is inversely proportional to the central charge $c$ of the theory so the result translates into a lower bound for $c$. 

\begin{figure}[H]
             \begin{center}             
              \includegraphics[scale=0.6]{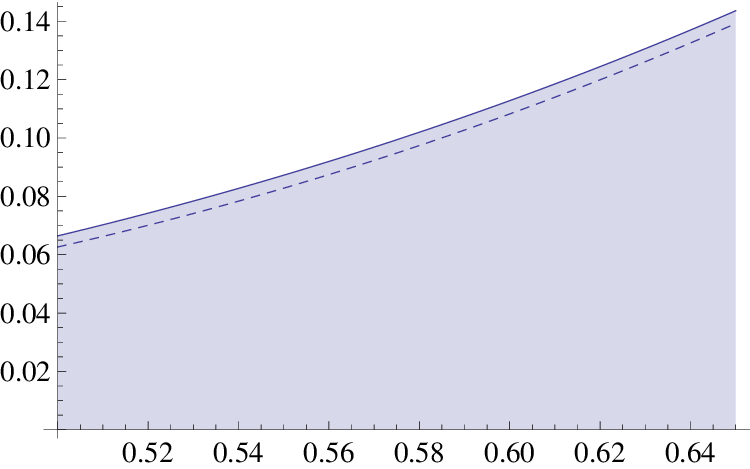}  
              \put(2,-5){$\Delta_{\text{ext}}$}    
              \put(-232,128){$\mu^2_d$}                          
              \caption{Upper bound for the coefficient of the $\hat T_{d d}$ block as a function of the external dimension. The dashed line represents and improved bound with a stronger assumption for the gap, following the dashed line of figure \ref{bulkextra} (see text).}           
              \label{upperbound}
            \end{center}
\end{figure}

In this section we will use the boundary bootstrap to bound the coefficient $\mu_d^2$ of the $\hat T_{dd}$ boundary block $f_{\text{bdy}}(d,\xi)$. We recall that this block is always present in the extraordinary transition, see the OPE in table \ref{tab:critexpising}. We start by imposing,
\bea
\Lambda(\xi^{\Delta_{\text{ext}}}f_{\text{bdy}}(d,\xi)) & = &  1\, ,
\\
\Lambda(F_\D) & \ge &  0\, .
\eea
Applying this functional to the crossing symmetry relation \eqref{sumrule} we obtain,
\be 
\mu_d^2 \le \Lambda(1)\, ,
\ee
where $\mu_d^2$ is the OPE coefficient of $f_{\text{bdy}}(d,\xi)$. The best bound is obtained by minimizing the action of $\Lambda$ on the identity. For the spectrum we require,
\be
\begin{split}
\Delta_{\text{bulk}} & \ge 2 \Delta_{\text{ext}}\, ,\\
\Delta^{(1)}_{\text{bdy}} &= 0\, ,\\
\Delta^{(2)}_{\text{bdy}} & \ge d\, .\\
\end{split}
\ee 
Notice that we have again assumed a gap of $2 \Delta_{\text{ext}}$ in the bulk. We plot our result as a function of the external dimension in figure \ref{upperbound}.
Let us try to justify our choice of $\Delta_{\text{bulk}} \ge 2 \Delta_{\text{ext}}$. A way to make the bound stronger would be to increase the bulk gap above this value, the maximum value we can assume for the gap is dictated by the bulk bound of \cite{ElShowk:2012ht}, obtained using the four-point bootstrap equations. In figure \ref{upperbound} we have thus plotted an improved upper bound (dashed line) assuming $\Delta_{\text{bulk}} \ge f(\Delta_{\text{ext}})$, where $f(\Delta)$ is the function represented by the dashed line of figure \ref{bulkextra}. It is clear that the upper bound is not too sensitive to the assumed gap. For example, for the Ising model $\Delta_{\text{ext}}=0.518$, and the upper bounds are $\mu_d^2 \lesssim 0.0734$ and $\mu_d^2 \lesssim 0.0693$ for $\Delta_{\text{bulk}} \ge 2(0.518) \sim 1.04$ and $\Delta_{\text{bulk}} \ge f(0.518)=1.41$ respectively. A change of $\sim 0.37$ in the bulk gap translates into a change of $\sim 0.0041$ in the bound, so at least for this example $2\Delta_{\text{ext}}$ does a good job as a representative gap for the space of CFTs.

The procedure used above generalizes with no major changes to arbitrary dimensions, let us then make a quick comparison with some known values. For the $2d$ Ising model the coefficient $\mu_d^2$ can be read from the conformal block expansion in (\ref{isingssbdy}), it has the value $\mu_d^2=\frac{1}{32\sqrt{2}} \sim 0.0221$ whereas the Linear Programming methods result in an upper bound $\mu_d^2 \lesssim 0.0309$. For the extraordinary transition in the $\epsilon$-expansion equation (\ref{epsextrabdy}) tells us $\mu_d^2=\frac{1}{10} = 0.10$, whereas we obtained the upper bound $\mu_d^2 \lesssim 0.119$ in four dimensions. We see that the numbers agree reasonably well.

\subsubsection{Towards the Ising model}
\begin{figure}[h!t]
\includegraphics[scale=0.6]{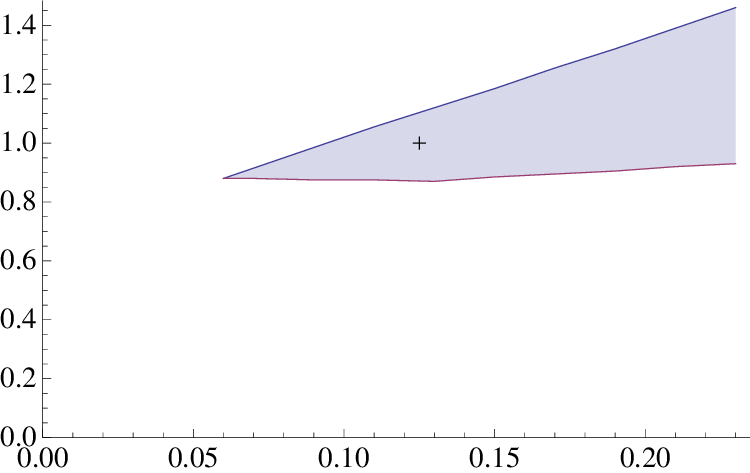}
\hfill 
\includegraphics[scale=0.6]{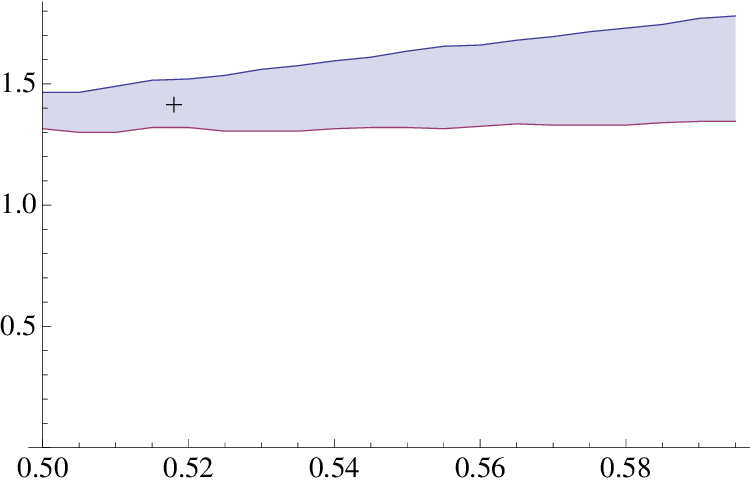}
\put(-500,130){$\Delta^{(1)}_{\text{bulk}}$}
\put(-235,130){$\Delta^{(1)}_{\text{bulk}}$}
\put(-255,-1){$\Delta_{\text{ext}}$}
\put(0,-1){$\Delta_{\text{ext}}$}
\caption{\label{fig:isingsearch}Locating the Ising model in $d=2$ (left) and $d=3$ (right). The plot show the dimension of a bulk operator versus the external dimension. With the assumptions explained in the main text, we need at least one bulk operator in the shaded regions. The Ising model is indicated with the cross in both plots.}
\end{figure}

In analogy with \cite{ElShowk:2012ht} we may try to isolate the Ising model in various dimensions. To this end we will improve the results of the previous subsection by using as additional knowledge the dimension of the next scalar operator $\veps'$ which appears in the $\s \times \s$ OPE. According to table \ref{tab:critexpising}, in three dimensions this operator has a scaling dimension $\D_{\eps'}$ of approximately 3.84 whereas in two dimensions it has dimension 4 (it corresponds to $L_{-2} \bar L_{-2} \mathbf 1$). We again assumed a boundary channel spectrum consistent with the extraordinary transition, \ie a possible one-point function and a gap equal to the spacetime dimensions $d$. Summarizing, 
\be
\begin{split}
\Delta^{(2)}_{\text{bulk}} & \ge \Delta_{\e'}\, ,\\
\Delta^{(1)}_{\text{bdy}} &= 0\, ,\\
\Delta^{(2)}_{\text{bdy}} & \ge d\, .\\
\end{split}
\ee
with $\D_{\eps'}$ fixed to the values of table \ref{tab:critexpising}. Our aim is now to find the possible range of values that $\Delta^{(1)}_{\text{bulk}}$ can take. The resulting plots are shown in figure \ref{fig:isingsearch}.

Notice that the plots give results that are qualitatively similar to those of \cite{ElShowk:2012ht}, in a considerably simpler setup. This is of course an encouraging result. Furthermore, we also did not rule out the Ising model and this provides some  \emph{a posteriori} justification for our assumption of positivity in three dimensions.

It is however rather unfortunate that the bounds we obtain are relatively weak. For this specific example we have tried different numerical implementations as well, for example we have tried to include more derivatives or to evaluate the blocks at different points like $\xi = 1/2$ or $\xi = 2$. In each case we were unable to significantly lower the bounds. We have also attempted to improve the results by imposing an additional gap between the second and the \emph{third} operator in the bulk channel. The third bulk operator has scaling dimensions $8$ in $d=2$  and approximately $4.6$ in $d=3$.  Imposing this additional gap significantly improved the bounds for $d=2$ but unfortunately this was not the case for $d=3$.

\section{Boundary crossing symmetry for stress tensors}
\label{sec:crossingsyenmom}
In section \ref{sec:bdycrossingsym} we derived the crossing symmetry equation \eqref{crossingsy} for two-point functions of scalar operators using the bulk and boundary conformal block decompositions. In this section we will derive a similar equation for the two-point function of the stress tensor. We will then use this equation in section \ref{sec:Tmnbounds} to obtain numerical bounds for the spectrum of operators appearing in the stress tensor OPE. 

The main results of this section are summarized  in subsection \ref{subsec:Tmnsummary}. We then present the details of our computations in subsections \ref{subsec:corrtensor} through \ref{subsec:boundaryblocksTmn}. These latter subsections are not essential for the remainder of the paper and can safely be skipped by the casual reader.

\subsection{Summary of results}
\label{subsec:Tmnsummary}
As we show in equation \eqref{tensorcorrelatorsbdycft} below, the two-point function of a spin two operator in the presence of a boundary features three independent tensor structures. Each tensor structure comes multiplied with its own scalar function of $\xi$ and we find it convenient to collect these three functions in a three-component vector of the form $\left(f(\xi),\,g(\xi),\,h(\xi)\right)$. Furthermore, for the stress tensor the Ward identities relate the three components in the following way:
\be
\label{wardidsTmn}
\begin{split}
(d-2) \xi^2 \frac{d}{d\xi}g &= (d^2 +3d - 2) h  - 2 (d-1) \xi(1+\xi) \frac{d}{d\xi}h\\
4 d \xi^3 \frac{d}{d\xi} f &= - 4 (1 + \xi) h + \Big(\xi(d^2 + 2d - 4) - 2d \xi^2(1+\xi) \frac{d}{d\xi} \Big) g\, ,
\end{split}
\ee 
so up to a few integration constants there is effectively only one independent function of $\xi$.

In the following subsections we derive the conformal block decompositions of the functions $(f,g,h)$ in the bulk and the boundary channel. The main result of these subsections will be the following crossing symmetry equation: 
\be
\label{crossingsyTmn}
\begin{split}
&\begin{pmatrix}
1 \\0 \\0
\end{pmatrix}
+ 
\sum_{k} \lambda_k a_{\op_k}
\begin{pmatrix}
f_{\text{bulk}}(\D_k;\xi)\\ g_{\text{bulk}}(\D_k;\xi)\\ h_{\text{bulk}}(\D_k;\xi)
\end{pmatrix}
\\
&\qquad \qquad = 
\mu_{(0)}^2
\begin{pmatrix}
f_\text{bdy}^{(0)}(d;\xi)\\ g_\text{bdy}^{(0)}(d;\xi)\\ h_\text{bdy}^{(0)}(d;\xi)
\end{pmatrix}
+
\mu_{(1)}^2
\begin{pmatrix}
f_\text{bdy}^{(1)}(d;\xi)\\ g_\text{bdy}^{(1)}(d;\xi)\\ h_\text{bdy}^{(1)}(d;\xi)
\end{pmatrix}
+
\sum_{n} \mu_{(2),n}^2
\begin{pmatrix}
f_{\text{bdy}}^{(2)}(\D_n;\xi)\\ g_{\text{bdy}}^{(2)}(\D_n;\xi)\\ h_{\text{bdy}}^{(2)}(\D_n;\xi)
\end{pmatrix}\, ,
\end{split} 
\ee
where all the functions $(f,g,h)$ are explicitly known functions of $\xi$. Equation \eqref{crossingsyTmn} is the analogue of \eqref{crossingsy} for scalars and we will use it in section \ref{sec:Tmnbounds} to obtain bounds on operator dimensions and OPE coefficients. Let us now discuss it in a bit more detail. 

First of all, because of the three independent tensor structures we get a three-dimensional vector of equations (and the conformal blocks themselves also become three-dimensional vectors). It is then important to realize that the Ward identities are operator equations and therefore they must be true for the individual conformal blocks as well. Each vector  appearing in \eqref{crossingsyTmn} thus \emph{individually} satisfies the Ward identities \eqref{wardidsTmn}.

The left-hand side of \eqref{wardidsTmn} is the bulk channel conformal block decomposition. As in \eqref{crossingsy}, we separated out the conformal block corresponding to the identity operator. For the other operators we should recall that $SO(d,1)$ conformal symmetry dictates that only scalars can get non-zero one-point functions and therefore only scalar blocks can contribute to the bulk channel expansion.

The right-hand side of \eqref{wardidsTmn} represents the boundary channel conformal block decomposition. A priori, a spin 2 operator has a boundary OPE decomposition involving operators with spins ranging from $0$ to $2$ and indeed we find all these possibilities in \eqref{wardidsTmn}, where the spins of the exchanged operator is written as the superscript in parentheses. However in this case the Ward identities turn out to further constrain the conformal block decomposition. More specifically, the boundary scalar and vector appearing in the boundary OPE decomposition of $T_{\m \n}$ \emph{must} have scaling dimensions equal to the spacetime dimensions, so $\Delta^{(0)} = \Delta^{(1)} = d$. There is thus a unique block for the exchange of a scalar of dimension $d$ and also for a vector of dimension $d$. These two blocks are the first two terms on the right-hand side of \eqref{crossingsyTmn}. On the other hand, the dimensions of the spin $2$ fields are not constrained in this way and there can therefore in principle be infinitely many spin 2 blocks, represented by the final sum in \eqref{crossingsyTmn}.

Let us offer a few more comments on the spin 0 and 1 boundary operators. As one may have anticipated, in physical theories they correspond to the $\hat{T}_{d\,d}$ and $\hat{T}_{i\,d}$ ($i$ being a tangential index)  components of the bulk stress tensor, restricted to the boundary. These operators are intimately related to infinitesimal variations in the location of the boundary surface which explains the `non-renormalization' of their scaling dimensions, see \cite{McAvity:1993ue} for details. For physical BCFTs the displacement operator $\hat{T}_{d\,d}$ is generically present on the boundary and we encountered it already in the discussion of the extraordinary transition in section \ref{sec:bootstrapinepsexpansion}. On the other hand, the vector operator is only present if there is a non-zero energy flow across the boundary. For BCFTs this is an unphysical boundary condition and we can then set $\mu_{(1)}^2 = 0$. (Notice that an energy flow would be allowed if the surface $x^d = 0$ was actually an $SO(d,1)$ preserving \emph{interface} between two different theories, one defined for $x^d > 0$ and the other for $x^d < 0$, and in such cases the vector block will generically be present.)

In appendix \ref{app:enmomtencorrs} we present a few explicit solutions to the crossing symmetry equation \eqref{crossingsyTmn}. We discuss the universal solution in two dimensions (which is fully determined by the Virasoro algebra), the free-field theory solutions in $d$ dimensions and the extraordinary transition to leading order for the Wilson-Fisher fixed point.

\subsection{Correlation functions of tensor operators}
\label{subsec:corrtensor}
In this section we discuss correlation functions of operators with spin in conformal field theories. We will use the results of \cite{Costa:2011mg}, see also \cite{Costa:2011dw}, and adapt them to conformal field theories with a boundary. Many of the results in this and the next two subsections were also obtained in \cite{McAvity:1995zd,McAvity:1993ue} but we present here an independent derivation which is straightforwardly implemented on a  computer.

The index structures appearing in correlation functions of tensor operators are easily found in the null projective cone formalism discussed in section \ref{subsec:scalar2ptfunc}. According to \cite{Costa:2011mg}, a generic tensor field $f_{\mu_1 \ldots \nu_n} (x)$ lifts to a tensor field $F_{A_1\ldots A_n}(P)$ in the null projective cone with the following properties:
\begin{itemize}
  \item equal symmetries in the indices of $F_{A_1\ldots A_n}(P)$ and of $f_{\mu_1 \ldots \nu_n} (x)$;
  \item transversality, so $P^{A_i} F_{A_1\ldots A_i \ldots A_n}(P) = 0$ for $1 \leq i \leq n$; 
  \item a gauge equivalence defined as $F_{A_1 \ldots A_n}(P) \sim F_{A_1 \ldots A_n}(P) + P_{A_i} \Lambda_{A_1 \ldots \hat A_i \ldots A_n}$ for any $\Lambda$ and $1 \leq i \leq n$.
\end{itemize}
For symmetric traceless tensors it is convenient to contract the indices on $F$ with auxiliary variables $Z^A$ and write $F(P,Z) \equiv F_{A_1 \ldots A_n}(P) Z^{A_1} \ldots Z^{A_n}$. Tracelessness implies that we may restrict ourselves to the subspace defined by $Z^2 = 0$ and the gauge equivalence implies that we may take $Z \cdot P = 0$ as well. The transversality condition becomes:
\be
P \cdot \frac{\del}{\del Z}\, F(P,Z) = 0\,.
\ee  

Correlation functions of $n$ symmetric traceless tensor primary operators can now be written as scalar functions $G(P_i,Z_i)$ with $1 \leq i \leq n$ with the following properties:
\begin{itemize}
\item the dependence on $Z_i$ should be a homogeneous polynomial of degree $l_i$;
\item the dependence on $P_i$ should be homogeneous of degree $- \D_i$;
\item transversality dictates that $P_i \cdot \del_{Z_i} G = 0$ for $1 \leq i \leq n$;
\item for any conserved tensor there is a Ward identity of the form \cite{Costa:2011mg}
 \be
(\del_{P} \cdot D^{(d)}) \,G = 0\, ,
 \ee
with
\be
\label{delzop}
D^{(d)}_A = \Big( \frac{d}{2} - 1 + Z \cdot \frac{\del}{\del Z}\Big) \frac{\del}{\del Z^A} - \hf Z_A \frac{\del^2}{\del Z \cdot \del Z}\, ,
\ee
where $P$ and $Z$ are the variables corresponding to the conserved tensor, for example $P_1$ and $Z_1$ if the conserved tensor is the first operator.
\end{itemize}

As an example, let us review the well-known result for the three-point function of two stress tensors and one scalar operator $G_{TT\op}(P_1,P_2,P_3,Z_1,Z_2)$. The first three constraints together dictate that there are three different invariant tensor structures,
\be
\label{GTTop}
G_{TT\op} = \frac{1}{(- 2 P_1 \cdot P_2)^{d-\D/2} (- 2 P_2 \cdot P_3)^{\D/2} (- 2 P_3 \cdot P_1)^{\D/2}} \Big(a (W_{12})^2 + b H_{12}^2 + c H_{12} W_{12} \Big)\,,
\ee
with for now arbitrary constants $a,b,c$ and with building blocks
\be
\label{H12W12}
\begin{split}
W_{12} &= \frac{\Big( (Z_1 \cdot P_2) (P_1 \cdot P_3) - (Z_1 \cdot P_3)(P_1 \cdot P_2) \Big) \Big( (Z_2 \cdot P_1) (P_2 \cdot P_3) - (Z_2 \cdot P_3)(P_1 \cdot P_2) \Big)}{(P_1\cdot P_2) (P_2 \cdot P_3) (P_3 \cdot P_1)}\, ,
\\
H_{12} &= \frac{(Z_1 \cdot Z_2)(P_1 \cdot P_2) - (Z_1 \cdot P_2)(Z_2 \cdot P_1)}{P_1 \cdot P_2}\,.
\end{split}
\ee
The Ward identities for the stress tensor furthermore dictate that:
\be
\label{abc}
\begin{split}
a &= \frac{\D (\D+2)}{4 d (d+1)} \lambda_{TT\op}\, , \\
b &= \frac{ (\D-d)^2 (d-1) - 2d}{d (d+1)(d-2)} \lambda_{TT\op}\, ,\\
c &= \frac{\D ((\D-d)(d-1) -2)}{d(d+1)(d-2)} \lambda_{TT\op}\, ,
\end{split}
\ee
where $\lambda_{TT\op}$ is an undetermined overall coefficient. Upon sending $\Delta \to 0$ we find that $a,c \to 0$ but $b \to \lambda_{TT\mathbf 1}$ and we recover the unit normalized stress tensor two-point function,
\be
\vev{T(P_1,Z_1) T(P_2,Z_2)} = \frac{H_{12}^2}{(-2 P_1 \cdot P_2)^{d}}\,,
\ee
provided we set $\lambda_{TT\mathbf 1} = 1$. The normalization in \eqref{abc} is therefore such that $\lambda_{TT\op}$ is a natural three-point coupling coefficient.

Let us finally take the OPE limit by sending $P_1 \to P_2$. In that case $H_{12}$ remains finite whilst
\be
\label{W12ope}
W_{12} \to W_{12}^{\text{OPE}} \equiv \frac{(Z_1 \cdot P_2) (Z_2 \cdot P_1)}{(P_1 \cdot P_2)}
\ee
and therefore
\be
G_{TT\op} \to \frac{  a (W_{12}^{\text{OPE}})^2 + b H_{12}^2  + c H_{12}W_{12}^{\text{OPE}}}{(- 2 P_1 \cdot P_2)^{d-\D/2} (- 2 P_1 \cdot P_3)^{\D} }\, ,
\ee
and we infer that the $T \times T \to \op$ operator product expansion becomes to leading order
\be
\label{TTOope}
T(P_1,Z_1) T(P_2,Z_2) \sim \ldots + \frac{a (W_{12}^{\text{OPE}})^2 + b H_{12}^2  + c H_{12}W_{12}^{\text{OPE}}}{(- 2 P_1 \cdot P_2)^{d-\D/2}} \op(P_1) + \ldots
\ee
where we assumed that $\op$ is normalized such that $\vev{\op(P_1) \op(P_2)} = (-2 P_1 \cdot P_2)^{-\Delta}$. 

As we mentioned in section \ref{subsec:scalar2ptfunc}, the breaking of $SO(d+1,1)$ to $SO(d,1)$ due to the presence of a boundary is implemented by introducing an additional fixed vector
\be
V^A = (0,0,\ldots,0,1)\,,
\ee
representing the unit normal to the boundary. Correlation functions are still required to be $SO(d+1,1)$ scalars with the same four properties as above but they can now depend on $V^A$ as well. For example, we have already mentioned that the one-point function of a scalar operator can take the form:
\be
\label{vevO}
\vev{\op(P)} = \frac{a_\op}{(V \cdot P)^\Delta}\, ,
\ee
with arbitrary coefficient $a_\op$. For one-point functions of tensor operators one directly sees that the numerator would have to involve a factor $(V \cdot Z)^l$ but this is not transverse and so higher-spin one-point functions must vanish.

With two points we can build the invariant object $\xi$ of section \ref{subsec:scalar2ptfunc} which we recall was
\be
\xi = \frac{- P_1 \cdot P_2}{ 2 (V \cdot P_1) (V \cdot P_2)} = \frac{(x_1-x_2)^2}{4 x_1^d x_2^d} \, ,
\ee
and conformal symmetry thus determines two-point functions only up to arbitrary functions of $\xi$. For the scalar two-point function this leads to equation \eqref{twoptscalar} which was:
\be
\vev{\op_1(P_1)\op_2(P_2)} = \frac{1}{(2 V \cdot P_1)^{\Delta_1} (2 V \cdot P_2)^{\Delta_2}} f_{\op_1\op_2}(\xi)\, ,
\ee
where $f_{\op_1\op_2}(\xi)$ is not fixed by conformal symmetry. Two-point functions involving tensors are easily found, e.g. 
\be
\label{tensorcorrelatorsbdycft}
\begin{split}
Z_2^A \vev{\op(P_1) \Jm_A(P_2)} &= \frac{(Z_2 \cdot V) (P_2 \cdot P_1) - (P_2 \cdot V) (Z_2 \cdot P_1)}{(V \cdot P_1)^{\Delta_\op + 1} (V \cdot P_2)^{\Delta_\Jm + 1}} f_{\op \Jm}(\xi)\, ,
\\
Z_2^A Z_2^B \vev{\op(P_1) \Tm_{AB} (P_2)} &= \frac{\Big((Z_2 \cdot V) (P_2 \cdot P_1) - (P_2 \cdot V) (Z_2 \cdot P_1)\Big)^2}{(V \cdot P_1)^{\Delta_\op + 2} (V \cdot P_2)^{\Delta_\Tm + 2}} f_{\op \Tm}(\xi)\, ,
\\
Z_1^A Z_2^B \vev{\Jm_A(P_1) \Jm_B(P_2)} &= \frac{f_{\Jm\Jm}(\xi) H_{12} + g_{\Jm\Jm}(\xi) Q_{12}}{\xi^{\Delta_1}(V \cdot P_1)^{\Delta_1} (V\cdot P_2)^{\Delta_2}}\, ,
\\
Z_1^A Z_1^B Z_2^C Z_2^D \vev{\Tm_{AB}(P_1) \Tm_{CD}(P_2)} &= \frac{f_{\Tm\Tm} (\xi) H_{12}^2 + g_{\Tm\Tm}(\xi) H_{12} Q_{12} + h_{\Tm\Tm}(\xi) Q_{12}^2}{(4\xi)^{\Delta_1} (V \cdot P_1)^{\Delta_1}(V \cdot P_2)^{\Delta_2}}\, ,
\end{split}  
\ee
with $H_{12}$ already defined above and with
\be
\label{Q12}
\begin{split}
Q_{12} = \left( \frac{(V \cdot P_1)(Z_1 \cdot P_2)}{(P_1 \cdot P_2)} - (V\cdot Z_1)\right) \left( \frac{(V \cdot P_2)(Z_2 \cdot P_1)}{(P_1 \cdot P_2)} - (V\cdot Z_2)\right)\, .
\end{split}
\ee
If the above tensors are conserved then we write $J$ and $T$ instead of $\Jm$ and $\Tm$. In that case $\Delta_J = d-1$ and $\Delta_T = d$ and from the Ward identities we also find that:
\be
\label{wardidsbdycorr}
\begin{split}
f_{\op J}(\xi) &= c_{\op J} (\xi (1+\xi))^{-d/2}\, ,
\\
f_{\op T}(\xi) &= c_{\op T} (\xi(1+\xi))^{-1-d/2}\, ,
\\
0 &=  \left((d+1) - 2 \xi \frac{d}{d \xi}\right) g_{JJ} - 2 \xi^2 \frac{d}{d\xi}  \Big(f_{JJ} + g_{JJ}\Big)\, ,
\\
(d-2) \xi^2 g'_{TT} &= (d^2 +3d - 2) h_{TT}  - 2 (d-1) \xi(1+\xi) h_{TT}'\, ,
\\
4 d \xi^3  f'_{TT} &= - 4 (1 + \xi) h_{TT} + \Big(\xi(d^2 + 2d - 4) - 2d \xi^2(1+\xi) \frac{d}{d\xi} \Big) g_{TT}\, ,
\end{split}
\ee 
with $c_{\ldots}$ denoting an integration constant. We see that the two-point function of two stress tensors and the two-point function of two currents are both fixed up to a single function of $\xi$. The last two equations in \eqref{wardidsbdycorr} were already presented in equation \eqref{wardidsTmn}. They agree with equation (2.27) and (2.31) of \cite{McAvity:1995zd} with the replacements $f(\xi) = C(v)$, $g(\xi) = 4v^2 B(v)$ and $h(\xi) = v^4 A(v)$ and with $v^2 = \xi / (\xi + 1)$.

We can also insert operators at boundary points labelled $X$ satisfying $X\cdot V = 0$. As before, we will denote such operators with a hat. We project the indices of such operators to lie along the boundary, which in the null projective cone is implemented by the constraint $V \cdot D^{(d)} = 0$ with the operator $D^{(d)}_A$ already given by \eqref{delzop}. The correlation functions of interest are those with a single stress tensor in the bulk. We find:
\be
\label{twoptTbulktobdy}
\begin{split}
Z_2^A Z_2^B \vev{\hat \op(X_1) T_{AB}(P_2)}&= \delta_{d,\Delta_{\hat \op}}\,\, c_{\hat \op T}  \frac{\Big((Z_2 \cdot V) (P_2 \cdot X_1) - (P_2 \cdot V) (Z_2 \cdot X_1)\Big)^2}{(-2 X_1 \cdot P_2)^{d + 2}}\, ,
\\
Z_1^A Z_2^B Z_2^C \vev{\hat \Jm_A (X_1) T_{BC}(P_2)}&= \delta_{d,\Delta_{\hat \Jm}}\,\, c_{\hat \Jm T}  \frac{\Big((Z_2 \cdot V) (P_2 \cdot X_1) - (P_2 \cdot V) (Z_2 \cdot X_1)\Big)  \hat H_{12}}{(- 2X_1 \cdot P_2)^{d + 1}}\, ,
\\
Z_1^A Z_1^B Z_2^C Z_2^D \vev{\hat \Tm_{AB}(X_1) T_{CD}(P_2)} &= c_{\hat \Tm T} \frac{\hat H_{12}^2 - \frac{1}{d-1} Q_{12}^2}{(- 2X_1 \cdot P_2)^{\Delta_{\hat \Tm}}(V \cdot P_2)^{d- \Delta_{\hat \Tm}}}\, ,
\end{split}
\ee
with
\be
\begin{split}
\hat H_{12} &= \frac{(\hat Z_1 \cdot Z_2)(P_1 \cdot P_2) - (\hat Z_1 \cdot P_2)(Z_2 \cdot P_1)}{P_1 \cdot P_2}\, , \qquad \qquad \hat Z_1^A \equiv Z_1^A - (Z_1\cdot V) V^A\,.
\end{split}
\ee
Notice that for scalars and vectors the scaling dimension is required to be $d$ whereas the dimension of $\hat \Tm$ is unconstrained by the Ward identity.

Up to terms that ensure that $V \cdot D^{(d)}$ annihilates the correlator, two-point functions of boundary operators are of the same form as two-point functions of bulk operators in the absence of a boundary. In particular we find that:
\be
\label{twoptbdy}
\begin{split}
\vev{\hat \op (X_1,Z_1) \hat \op (X_2,Z_2)} &=  \frac{1}{(-2X_1 \cdot X_2)^{\Delta}}\, ,
\\
\vev{\hat \Jm (X_1,Z_1) \hat \Jm (X_2,Z_2)} &=  \frac{H_{12} - (V\cdot Z_1)(V \cdot Z_2)}{(-2X_1 \cdot X_2)^{\Delta}}\, ,
\\
\vev{\hat \Tm (X_1,Z_1) \hat \Tm (X_2,Z_2)} &=  \frac{\Big(H_{12}- (V\cdot Z_1)(V \cdot Z_2)\Big)^2 - \frac{1}{d-1}(V \cdot Z_1)^2 (V \cdot Z_2)^2}{(-2X_1 \cdot X_2)^{\Delta}}\,.
\end{split}
\ee
Equation \eqref{twoptbdy} defines our normalization of the boundary operators. Notice that $H_{12}$ descends from the projective cone to $z_{1}^\m z_{2}^\n (\delta_{\m \n} - 2 x_{12,\m} x_{12,\n} / x_{12}^{2})$ so it is easily verified that our normalization is consistent with reflection positivity. Using \eqref{twoptTbulktobdy} and \eqref{twoptbdy} we find the bulk-to-boundary OPE of the stress tensor,
\be
\label{TbdyOPE}
T(P,Z) \to c_{\hat \op T} (Z\cdot V)^2 \hat \op(X) - c_{\hat \Jm T} (Z \cdot V) \hat \Jm (X,Z) + \frac{c_{\hat \Tm T}}{(V \cdot P)^{d - \Delta_{\hat T}}} \hat \Tm (X,Z) + \ldots
\ee

\subsection{Bulk channel blocks for the stress tensor}
In this subsection we compute the conformal blocks for the two-point function of the stress tensor using the conformal Casimir differential equation method of \cite{Dolan:2003hv}. These are the conformal blocks appearing on the left-hand side of \eqref{crossingsyTmn}.

On a symmetric traceless tensor $F(P,Z)$ the action of an element $L_{AB}$ of $SO(d+1,1)$ takes the form:
\be
L_{AB} F(P,Z) = \Big( P_A \frac{\del}{\del P^B} - P_B \frac{\del}{\del P^A} + \frac{1}{\frac{d}{2} + l - 1} (Z_A D^{(d+2)}_B - Z_B D^{(d+2)}_A) \Big) F(P,Z)\, ,
\ee
with the operator $D^{(d+2)}_A$ given by \eqref{delzop} but with $d \to d+2$ since we are rotating in $d+2$ dimensions. The conformal Casimir equation is then:
\be
\hf L_{AB}L^{AB} F(P,Z) = - C_{\Delta,l} F(P,Z)\, ,
\ee
with $C_{\Delta,l} = \Delta (\Delta - d) + l (l+d-2)$. We used this equation in appendix \ref{app:scalarblockders} to find the result \eqref{bulkblocktwoscalars} for the conformal block in the bulk channel for a scalar two-point function. For two stress tensors the conformal block can be written as:
\be
G_{b}^\Delta(P_1,P_2,Z_1,Z_2)  = \frac{f_{b} (\xi) H_{12}^2 + g_{b}(\xi) H_{12} Q_{12} + h_{b}(\xi) Q_{12}^2}{(4\xi)^d (V \cdot P_1)^{d}(V \cdot P_2)^{d}}\, ,
\ee
and the constraint $\hf (L^{(1)}_{AB} + L^{(2)}_{AB}) (L^{(1)AB} + L^{(2)AB})G_{}^\Delta = - C_{\Delta,0} G_{}^\Delta$ together with the Ward identities leads to the unique solution for the coefficients:
\be
\label{hbulk}
\begin{split}
h_{b} &=  \frac{\D(\D+2)}{16 d(d+1)}  \, (4\xi)^{\D/2 + 2} {}_2 F_1 \left(2 + \frac{\D}{2},2 + \frac{\D}{2}; 1 - \frac{d}{2} + \Delta; - \xi \right)\, ,\\
\end{split} 
\ee
with $f_{b}$ and $g_{b}$ determined by the Ward identities \eqref{wardidsbdycorr}. Let us verify the normalization by taking the OPE limit $\xi \to 0$. We already mentioned that $H_{12}$ then remains finite and it is not hard to find that
\be
Q_{12} \to - \frac{1}{2\xi} W_{12}^{\text{OPE}}\, ,
\ee
with $W_{12}^{\text{OPE}}$ defined in \eqref{W12ope}. From the expansion of \eqref{hbulk} and the Ward identities we find
\begin{align}
h_b &= (4\xi)^{\D/2} (  4 \xi^2 \hat a  + O(\xi))\, , && \hat a =  \frac{\D(\D+2)}{4 d(d+1)}\, , \nonumber\\
f_b &= (4\xi)^{\D/2} \Big(  \hat b + O(\xi) \Big)\, ,  && \hat b = \frac{(\D-d)^2(d-1) - 2 d}{d (d+1) (d - 2)}\, , \\
g_b &= (4 \xi)^{\D/2} \Big( -  2 \xi \hat c  + O(\xi)\, , \Big) && \hat c = \frac{\Delta ((\D - d)(d-1) - 2)}{d (d+1) (d-2)}\, ,  \nonumber
\end{align}
and the entire block behaves as:
\be
G_{b}^\Delta(P_1,P_2,Z_1,Z_2)  =  \frac{ \hat a (W_{12}^{\text{OPE}})^2 + \hat b H_{12}^2 + \hat c H_{12} W_{12}^{\text{OPE}}}{(- 2 P_1 \cdot P_2)^{d-\D/2}(V \cdot P_1)^\Delta}\, ,
\ee
which is compatible with \eqref{abc}, \eqref{TTOope} and \eqref{vevO}.

Explicit expressions for $f_b$ and $g_b$ are also available in terms of linear combinations of ${}_2 F_1$ hypergeometric functions.

The identity block can be found by sending $\Delta \to 0$. We then find that $f_b = 1$ and $g_b = h_b = 0$.

\subsection{Boundary channel blocks for the stress tensor}
\label{subsec:boundaryblocksTmn}
We label the boundary block associated to a primary operator of dimension $\Delta$ and spin $l$ as $G_s^{(\D,l)}$ (with a subscript ``s'' for surface). Each block has again the same form as the $TT$ two-point function given in \eqref{tensorcorrelatorsbdycft} with three associated functions $f_s^{(\D,l)}$, $g_s^{(\D,l)}$ and $h_s^{(\D,l)}$. In the two-point function of the stress tensor there are three types of boundary blocks, $G_s^{(d,0)}$, $G_s^{(d,1)}$ and $G_s^{(\Delta,2)}$. To find these blocks we act with the $SO(d,1)$ Casimir operator on one of the two points and solve the resulting differential equation. In the equations below we use $h \equiv d/2$.

For a block corresponding to the exchange of a boundary scalar of dimension $d$ we find:
\be
\begin{split}
h_s^{(d,0)} &= \frac{1}{ 2 h (2 h + 1)} \xi^{h + 1} (1+ \xi)^{-h-3}\Big( 2h (2h + 1) \xi^2 + 2(2h + 1 ) (h-1) \xi + h(h-1) \Big)\, ,
\\
g_s^{(d,0)} &= \frac{1}{h (2h+1)} \xi^{h}(1 + \xi)^{-h - 2}  (h + \xi + 2 h \xi)\, ,
\\
f_s^{(d,0)} &= \frac{1}{4 h (2h + 1)} \xi^{h - 1} (1+ \xi)^{-h -1}\, ,
\end{split}
\ee
where we already fixed the normalization. In the limit where $\xi \to \infty$ we find that only the third tensor structure contributes and
\be
G_s^{(d,0)}(P_1,P_2,Z_1,Z_2) \sim \frac{(V \cdot Z_1)^2 (V \cdot Z_2)^2}{(-2 P_1 \cdot P_2)^{2h}}\, ,
\ee
which agrees with \eqref{TbdyOPE} and the first equation in \eqref{twoptbdy}.

For the block corresponding to the exchange of a boundary vector of dimension $d$ we find:
\be
\begin{split}
h_s^{(d,1)} &= \frac{1}{2 (2h + 1)} \xi^{h+1} (1+\xi)^{-h-3} \Big(- 2 (2h + 1) \xi^2 + 2 h (h-1)\xi + h (h-1) \Big)\, ,
\\
g_s^{(d,1)} &= \frac{1}{(2h+1)} \xi^h (1+ \xi)^{-h-2} \Big( \xi^2 + h (1 + 2 \xi + 2 \xi^2) \Big)\, ,
\\
f_s^{(d,1)} &= \frac{1}{4 (2h + 1)} \xi^{h-1} (1+\xi)^{-h-1} (1 + 2 \xi)\, ,
\end{split}
\ee
and the block behaves for $\xi \to \infty$ as
\be
G_s^{(d,1)}(P_1,P_2,Z_1,Z_2) \sim \frac{(V \cdot Z_1) (V \cdot Z_2) \Big(H_{12} - (V \cdot Z_1) (V \cdot Z_2) \Big)}{(-2 P_1 \cdot P_2)^{2h}}\, ,
\ee
which is again consistent with the formulas given above.

Finally, for the spin two blocks:
\begin{align}
h_s^{(\D,2)} &= \frac{2 (h-1)}{2h-1} (4\xi)^{2h - \D} {}_3 F_2\Big( 2 + \D, 3 - 2h + \D ,1 - h + \D; 1 - 2h + \D, 2- 2h + 2\D ; - \frac{1}{\xi} \Big)\, , \nonumber\\
 g_s^{(\D,2)} &= -2 (4\xi)^{2h - \D} + O(\xi^{-1})\, ,\\
 f_s^{(\D,2)} &= (4\xi)^{2h - \D} + O(\xi^{-1})\, , \nonumber
\end{align}
where $g_s^{(\D,2)}$ and $f_s^{(\D,2)}$ can also be explicitly written as a sum over two hypergeometric functions. As we send $\xi \to \infty$ we recover that
\be
G_s^{(\D,2)}(P_1,P_2,Z_1,Z_2) \sim \frac{\Big(H_{12}- (V\cdot Z_1)(V \cdot Z_2)\Big)^2 - \frac{1}{d-1}(V \cdot Z_1)^2 (V \cdot Z_2)^2}{(V \cdot P_1)^{2h - \D} (V \cdot P_2)^{2h - \D} (- 2 P_1 \cdot P_2)^{\Delta}}\, ,
\ee
which is again consistent with the formulas given above.

\section{Numerical results for stress tensors}
\label{sec:Tmnbounds}
The numerical analysis of equation \eqref{crossingsyTmn} proceeds largely as for the scalar two-point function, see subsection \ref{subsec:scalarnumerimpl}. In particular, we again translate the constraints of crossing symmetry to an infinite vector of derivatives at $\xi = 1$ and apply a linear functional in order to exclude certain spectra, using the same numerical methods as described above. Notice that the Ward identities \eqref{wardidsTmn} can be used to express derivatives of $f$ and $g$ in terms of derivatives of $h$. We therefore do not need to include more than the zeroth derivative for the $f$ and $g$ components if we include many derivatives of the $h$ component. There is again no guarantee that the coefficients of the conformal blocks are positive in the bulk channel. Just as before we will therefore have to assume this condition of positivity in order to obtain any bounds.

\subsection{Bound on the bulk gap}
\begin{figure}[h!]
            \begin{center}
            \includegraphics[scale=0.6]{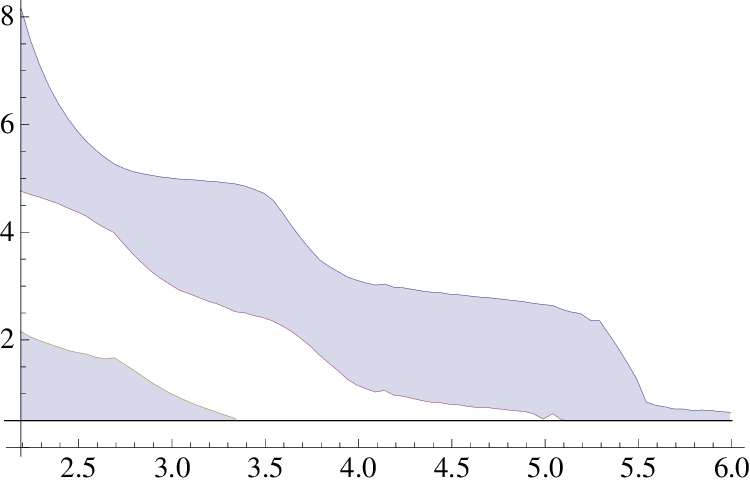}
               \put(-0,-8){$\D_{(2)}$}
               \put(-240,125){$\Delta_\text{bulk}$} 
             \caption{\label{fig:enmomten} Bounds for the energy momentum tensor two-point function in three spacetime dimensions. The upper curve is the upper bound $\D_\text{bulk}$ for the first bulk operator as a function of the gap $\D_{(2)}$ for the first spin 2 boundary operator. The other lines denote further constraints for such a bulk operator, to the extend that for every $\D_{(2)}$ there has to be at least one bulk scalar somewhere in the shaded region.}
            \end{center}
\end{figure}

In order to turn equation \eqref{crossingsyTmn} into a useful equation to constrain conformal field theories we have to decide which parameters we are going to vary. In previous computations of this sort the canonical parameter was always the dimension of the external field but for the stress tensor this dimension is of course fixed to be $d$. In our first analysis we instead chose to vary the dimension of the lowest spin $2$ boundary block which we denote as $\D_{(2)}$.  We then obtained an upper bound for the lowest bulk operator dimension as a function of $\D_{(2)}$ which we plotted as the upper curve in figure \ref{fig:enmomten}. 

We may rephrase this result by saying that the upper curve in figure \ref{fig:enmomten} informs us that the crossing symmetry equation \eqref{crossingsyTmn} can only be satisfied if there is at least one ``critical'' bulk operator with a scaling dimension somewhere below the curve. We can however subsequently ask whether this ``critical'' operator really could be sitting anywhere below the curve (and above the unitarity bound $\D_{\text{bulk}} > 1/2$). In fact it turns out that the region where such an operator has to appear can be constrained even further: we can limit it to the shaded region in figure \ref{fig:enmomten}. We conclude that for every $\D_{(2)}$ there has to be at least one bulk operator somewhere within this region. (There could in addition be other operators, for example somewhere in the white ``band'' or multiple operators in the shaded region, but none of this modifies the validity of our claim.)

In figure \ref{fig:enmomten} we assumed that the vector block was \emph{not} present in the boundary OPE of $T_{\m \n}$. Upon repeating the analysis with a vector block we obtained exactly the same curves for $\D_{(2)} > 3$ (up to small deviations due to the finite numerical precision), whereas for $\D_{(2)} \leq 3$ we would not be able to bound the bulk gap at all. The latter phenomenon has an easy explanation: the bulk identity operator can be decomposed in the boundary channel into the scalar block, the vector block and an infinite series of spin 2 blocks starting with $\D_{(2)} = 3$. For $\D_{(2)} \leq 3$ and with the vector block present it is therefore possible to have an infinite gap in the bulk (\ie no bulk operators apart from the identity) and so $\D_\text{bulk}$ cannot be bounded. This is reminiscent of the ``trivial'' solution for the scalar two-point function discussed in appendix \ref{subsec:trivialsol} which we found numerically in section \ref{subsec:special}.

The curves shown in figure \label{fig:enomten} have several ``bumps'' and other features whose origins are unfortunately unclear to us. For example, we were unable to find specific solutions of crossing symmetry that reflect the existence of these bumps. It would be interesting to see if such solutions exist and whether a conformal field theory is associated to them.

\subsection{Bound on OPE coefficients in the three-dimensional Ising model}
\begin{figure}[h!]
             \begin{center}        
            \includegraphics[scale=0.6]{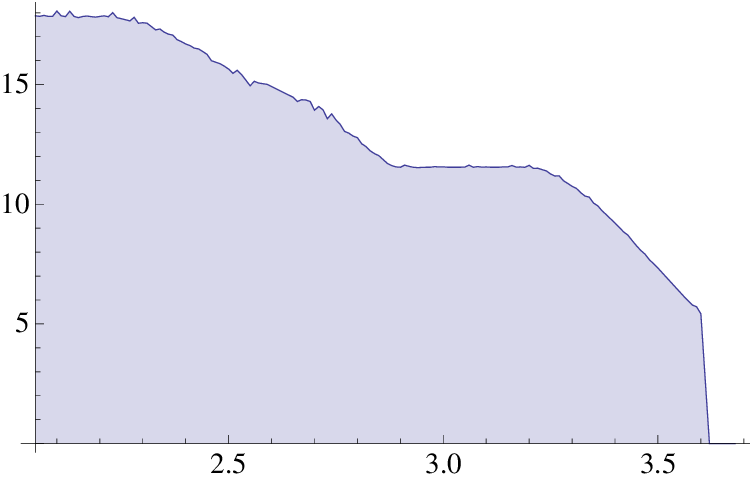}
               \put(-0,-8){$\D_{(2)}$}
               \put(-235,125){$\mu_{(0)}^2$} 
              \caption{\label{fig:enmomtenOPEcoeff} Bounds for the coefficient of the scalar boundary block in the two-point function of the stress tensor as a function of the gap $\D_{(2)}$ in the spin 2 boundary dimensions.}
            \end{center}
\end{figure}

In subsection \ref{subsubsec:boundopecoeffscalar} we discussed how to bound OPE coefficients in the conformal block decomposition. Here we repeat the same procedure for the two-point function of the stress tensor. We will again bound the coefficient of the boundary operator $\hat T_{d d}$ which in equation \eqref{crossingsyTmn} corresponds to the coefficient $\mu_{(0)}^2$ of the scalar block in the boundary channel. In addition, we decided to focus our attention on the three-dimensional Ising model. In particular, we have assumed that the bulk spectrum consists of operators with dimensions equal to $1.41$, $3.84$, and any operator with a scaling dimension greater than $4.6$. We then obtain an upper bound on $\mu_{(0)}^2$  as a function of the unknown scaling dimension $\D_{(2)}$ of the lowest spin two operator in the boundary channel. We assumed that no vector operator was present in the boundary channel. Our results are plotted in figure \ref{fig:enmomtenOPEcoeff}.

We find a rather surprising plateau for $\D_{(2)}$ between approximately 2.9 and 3.2 where $\mu_{(0)}^2 \sim 11.5$. From the results in appendix \ref{app:enmomtencorrs} we find that $\mu_{(0)}^2 = 4$ in two dimensions and that $\mu_{(0)}^2 = 640 / \epsilon + O(\e^0)$ in $4- \eps$ dimensions so at the very least our estimate appears to have the right order of magnitude. It would be interesting to compute the dimension of the first spin 2 operator appearing in the boundary channel in the epsilon expansion, since it is natural to expect that the Ising model lies at one of the corners of this plateau.

\section{Conclusions}

In this paper we have begun to explore the constraining power of crossing symmetry for BCFTs in general spacetime dimensions. After discussing the basic setup in section \ref{sec:bdycrossingsym}, we illustrated the relative simplicity of the ``boundary bootstrap'' in section \ref{sec:bootstrapinepsexpansion} where we found exact solutions with at most two blocks in each channel. We have then applied the linear programming methods of \cite{Rattazzi:2008pe} to the boundary crossing symmetry equations for
 both scalar operators and  stress tensors. With our assumption of positivity for the bulk expansion coefficients, we have demonstrated that these methods can be useful in the BCFT setup as well and that they lead to interesting universal bounds on scaling dimensions and OPE coefficients. Several of our results warrant a more detailed theoretical investigation. For example, the bound on the second boundary operator in the special transition and the $T_{d d}$ OPE coefficient in the extraordinary transition should be compared with computations in the epsilon expansion. Similarly, our numerical results of section \ref{subsec:special} indicate that the bulk-to-boundary OPE always has to be singular, a result that should be put on a more solid theoretical footing. Finally, our results for the stress tensor are rather mysterious and certainly call for further investigations, beginning with the one-loop anomalous dimension of the spin two boundary operator in the extraordinary transition.

It is unfortunate that the distinct ``kinks'' of \cite{ElShowk:2012ht} appear not to be generically present in the BCFT bounds. We emphasize that this (negative) result is completely independent from our positivity assumption, indeed in $d=2$ we see no kink but we know that the exact result does exhibit positivity. It would be interesting to see if there is another solution to crossing symmetry ``standing in the way'' and thereby preventing us from obtaining such a kink. More generally our results are a reflection of the fact that there is currently no 
deep understanding
of why and when such kinks will appear. It would of course be very interesting to understand this phenomenon better. We hope hat our numerical results (as well as the analytical results for the minimal models of appendix \ref{subsec:2dising}) will be helpful in further investigations.

The weakest point of our analysis is admittedly the assumption of positivity for the bulk expansion coefficients. 
While we have presented strong evidence that it is satisfied for the special and extraordinary Ising BCFTs,
it would be desirable to find a proof. A possible approach would be to derive
rigorous inequalties for boundary correlators on the lattice.

This paper is a first attempt to investigate the boundary bootstrap with a focus on the three-dimensional Ising model, but we feel we have just scratched the surface and that there are many interesting open questions. It is clear that the avenues for further numerical exploration are practically unlimited, but let us discuss a few possibilities in more detail.

First of all we could consider other scalar two-point functions to further investigate the spectrum of the three-dimensional Ising model. For example, one can try to further constrain the $\mathbb Z_2$ even scalar boundary spectrum by analyzing the two-point function of the $\varepsilon$ operator beyond what is shown in figure \ref{sndbdyoperatoree}. Of course this is straightforward: although in section 4 we mostly referred to the external operator as the $\sigma$ operator, in fact the bounds we obtained applied to any two-point function of identical scalar bulk operators, 
so the $\langle \varepsilon(x_1) \varepsilon(x_2)\rangle$ two-point function can be analyzed by simply dialing the external dimension to the right value
and relaxing the constraint from the $\mathbb{Z}_2$ selection rule. For the extraordinary transition one could also try to probe the $\mathbb Z_2$ odd one-point functions by studying a mixed two-point function like $\langle \sigma(x_1) \varepsilon (x_2) \rangle$.

Another class of options is to study correlation functions involving boundary operators. Here we find non-trivial structures in \emph{e.g.} the three-point function of two bulk operators and one boundary operator or the three-point function with two boundary operators and one bulk operator. However, in the former case it is clear that positivity cannot be guaranteed in either channel, whereas in the latter case there is only one conformal block decomposition so there is no crossing symmetry condition. These correlators could nevertheless be useful by conjecturing additional positivity constraints or by considering the constraints arising from multiple correlators at the same time.

Perhaps the most promising correlator 
is the four-point function of four boundary operators, which should lead to non-trivial constraints for the boundary spectrum. Here the positivity
assumption is certainly satisfied for any unitary boundary condition. The two-dimensional bounds of \cite{Rychkov:2009ij} also apply to the boundary
spectrum of a $3d$ theory, and we have checked for example
that the spectrum for the ordinary and special transition in the $3d$ Ising model (estimated from the epsilon expansion at one loop) lies strictly below the bounds. In fact
one should be able to do better: since the boundary spectrum does not involve a stress tensor one can additionally impose a finite gap (above the unitarity bound) for the first spin two operator. One can then study how  the upper bound on the dimension of the first scalar will come down if one increases this gap.
It will be very interesting to see ``kinks'' appear in such an analysis. We hope to report the results of this analysis in future work.

There are many more general directions to pursue as well. To mention a few, one may extend our results to supersymmetric theories, and to spacetime dimensions greater than four. Furthermore, the relatively simple form of the conformal blocks makes the boundary bootstrap especially suitable for investigations involving tensor operators, a research direction that is much more involved for the bulk four-point function in a theory with no boundary. Finally there is the prospect to broaden the setup and include conformal defects of all possible codimensions.

\section*{Acknowledgments}
It is a pleasure to thank all the participants  to the ``Back to the Bootstrap II'' workshop (PI, June 2012) for many exciting discussions.
We thank the Perimeter Institute for the hospitality and for providing the  inspiring environment
where this work took off. We are especially grateful
 to the authors of \cite{ElShowk:2012ht} for making available their numerical results, to
 Sheer El-Showk for sharing his CPLEX linear programming interface, and  to Slava Rychkov for very useful guidance to the linear programming methods. LR would also like to thank Riccardo Rattazzi and Alessandro Vichi for early discussions that deepened his interest in the bootstrap. We also thank
 the JHEP referee for suggesting some avenues for future research.
 This work is partially supported by the NSF under Grants PHY-0969919 and PHY-0969739. Any opinions, findings, and conclusions
or recommendations expressed in this material are those of the authors and do not necessarily
reflect the views of the National Science Foundation.

\appendix

\section{Scalar conformal blocks}
\label{app:scalarblockders}

In this section we will use the method of \cite{Dolan:2003hv} to obtain the scalar conformal blocks as eigenfunctions of the conformal Casimir operator. This procedure can be applied with no major changes to two-point functions involving tensor operators, and it was used succesfully in section \ref{sec:crossingsyenmom} to decompose the two-point function of the stress tensor.

\subsection*{Bulk channel}


The $SO(d+1,1)$ generators are,
\be 
L_{AB} = P_A \frac{\partial}{\partial P^B}-P_B \frac{\partial}{\partial P^A}\, ,
\ee
where $P^A=(P^+,P^-,P^1,\ldots P^{d})$.
To obtain the conformal blocks we solve the eigenvalue problem \cite{Dolan:2003hv},
\be 
\label{caseq}
L^2 \langle \cal O_1(P_1) \cal O_2(P_2) \rangle = -C_{\Delta,0} \langle \cal O_1(P_1) \cal O_2(P_2) \rangle\, ,
\ee
with $L^2=\frac{1}{2}(L^{(1)}_{AB}+L^{(2)}_{AB})(L^{(1)\, AB}+L^{(2)\, AB})$ and $C_{\Delta, l} = \Delta (\Delta - d)+l(l + d - 2)$, where $\Delta$ and $l$ are the dimension and spin of the internal operator. Because of Lorentz invariance no operators with spin can ever appear in the bulk conformal block decomposition, hence we set $l=0$ in equation (\ref{caseq}).

Once the asymptotic behavior of $f(\xi)$ is given, the conformal block is completely fixed. In the $\xi \rightarrow 0$ limit the bulk OPE dictates \cite{McAvity:1995zd},
\be 
\label{boundcon}
f(\xi) \sim \xi^{-\frac{1}{2}(\Delta_1+\Delta_2-\Delta)}\, .
\ee
where $\Delta_1$ and $\Delta_2$ are the dimensions of the external operators.
Stripping out this factor $f(\xi)=\xi^{-\frac{1}{2}(\Delta_1+\Delta_2-\Delta)} g(\xi)$ and plugging in (\ref{caseq}) we obtain a standard hypergeometric equation, 
\be 
\xi(1+\xi)g''(\xi)+(c+(a+b+1)\xi)g'(\xi)+a b g(\xi) = 0\, ,
\ee
with,
\be 
a = \frac{1}{2}(\Delta + \Delta_1-\Delta_2)\,, \quad b = \frac{1}{2}(\Delta - \Delta_1 + \Delta_2)\,, \quad c = \Delta -\frac{d}{2}+1\,.
\ee
The conformal block for the bulk channel is then,
\be 
\label{bulkblocktwoscalars}
f(\xi)=\xi^{-\frac{1}{2}(\Delta_1+\Delta_2-\Delta)} {_2}F_1\Big( \frac{1}{2}(\Delta + \Delta_1-\Delta_2),\frac{1}{2}(\Delta - \Delta_1 + \Delta_2),\Delta -\frac{d}{2}+1;-\xi\Big)\, ,
\ee
in perfect agreement with \cite{McAvity:1995zd}.

\subsection*{Boundary channel}

In this channel we consider the restricted conformal group. The $SO(d,1)$ generators are,
\be 
L_{ab} = P_a \frac{\partial}{\partial P^b}-P_b \frac{\partial}{\partial P^a}\, .
\ee
where $P^a=(P^+,P^-,P^1,\ldots P^{d-1})$.
To obtain the conformal blocks we act with the Casimir operator on one of the fields and solve the eigenvalue problem,
\be 
\label{2ptbnd}
L^2 \langle \cal O_1(P_1) \cal O_2(P_2) \rangle = -C_{\Delta,0} \langle \cal O_1(P_1) \cal O_2(P_2) \rangle\, .
\ee
where $C_{\Delta, l} = \Delta (\Delta - d+1)+l(l + d - 3)$ in this case.
For this particular two point function only scalar blocks are present, so $l=0$ again. However, this is no longer true for operators with indices (see subsection \ref{subsec:boundaryblocksTmn}). The asymptotic behavior for $\xi \rightarrow \infty$ can be obtained from the bulk-to-boundary OPE \cite{McAvity:1995zd},
\be 
\label{asym}
f(\xi) \sim \xi^{-\Delta}\, .
\ee
Stripping out this factor and plugging in (\ref{2ptbnd}) we obtain another hypergeometric equation.
The boundary block is, 
\be 
f(\xi)=\xi^{-\Delta} {_2}F_1\Big( \Delta,\Delta-\frac{d}{2}+1,2\Delta + 2 - d;-\frac{1}{\xi}\Big)\, ,
\ee
again in perfect agreement with \cite{McAvity:1995zd}.

\section{Solutions to crossing symmetry for scalar operators}
\label{app:crossingsysolns}
In this section we discuss a few solutions to the crossing symmetry equations for scalar two-point functions. We have:
\be
\vev{O(x_1) O(x_2)}= \frac{1}{(2x^d_1)^\Delta (2 x^d_2)^\Delta} \xi^{-\Delta} G(\xi)
\ee
where $\Delta$ is the conformal dimension of the operator $O$. The conformal block decomposition is,
\be
G(\xi) = 1 + \sum_{k} \lambda_k a_{k}f_{\text{bulk}}(\Delta_k;\xi) = \xi^{\Delta} \sum_{l} \mu_{l}^2 f_{\text{bdy}}(\Delta_l;\xi)
\ee
with $\lambda_k$ and $\mu_k$ three-point couplings and $a_{k}$ the coefficient of the one-point function of the $k$'th operator.

\subsection{Two-dimensional Ising model}
\label{subsec:2dising}
In this section we will decompose several correlators for the two-dimensional Ising model. The basic fields of the theory, corresponding to the energy and spin operators, will be denoted by $\veps$ and $\s$ respectively and have scaling dimensions $\D_{\veps} = 1$ and $\D_{\sigma} = \frac{1}{8}$ respectively. As we discussed in section \ref{sec:bootstrapinepsexpansion}, there are three different conformally invariant boundary conditions (or boundary states), given in equations \eqref{cardystatesextr} and \eqref{cardystatesord}. The first two are related by the $\mathbb Z_2$ symmetry of the theory and result in the same two-point function of $\sigma$.  The two remaining possible two-point functions for the $\sigma$ field are then \cite{DiFrancesco:1997nk},
\be
G_{\s \s}^{\pm} = \xi^{1/8}\sqrt{\Big(\frac{1+ \xi}{\xi}\Big)^{1/4} \pm \Big(\frac{\xi}{1+\xi}\Big)^{1/4}}\, . 
\ee
As we shall see below, the $+$ sign corresponds to the extraordinary transition, \emph{i.e.} the $| \mathbf 1 \rangle\! \rangle$ and $| \veps \rangle\! \rangle$ Cardy boundary states, whereas the $-$ sign corresponds to the ordinary transition which is the $| {\sigma} \rangle\! \rangle$ Cardy boundary state.

The full conformal block decomposition can in principle be obtained from Virasoro representation theory. We content ourselves here with a simpler analysis where we expand the correlation function in the limits $\xi \to 0$ and $\xi \to \infty$ and match the coefficients of the expansion to conformal blocks.
The bulk block decomposition becomes
\be
G_{\s \s}^{\pm} = 1 \pm \hf f_{\text{bulk}}(1;\xi) + \frac{1}{64} f_{\text{bulk}}(4;\xi) + \frac{9}{40960} f_{\text{bulk}}(8;\xi) \pm \frac{1}{32768} f_{\text{bulk}} (9,\xi) + \ldots
\ee  
The bulk spectrum corresponds to the identity $\mathbf 1$ and the energy $\veps$ operators plus scalar Virasoro descendants. For example, we may identify the operator of dimension $4$ with $L_{-2} \bar L_{-2} \mathbf 1$ and the operator of dimension $9$ with a level four descendant of $\veps$. The absence of an operator of dimension $5$ is in agreement with the fact that $\veps$ has a null descendant at level two, so $L_{-2} \bar L_{-2} \veps$ is actually an $SO(2,2)$ descendant.

In the boundary channel we find that:
\be
\label{isingssbdy}
\begin{split}
\xi^{-\Delta_{\s}} G_{\s \s}^+ &= \sqrt{2} + \frac{1}{32 \sqrt{2}} f_{\text{bdy}}(2;\xi) + \frac{9}{20480 \sqrt{2}} f_{\text{bdy}}(4;\xi) + \frac{25}{1835008 \sqrt{2}} f_{\text{bdy}}(6;\xi) + \ldots\\
\xi^{-\Delta_{\s}} G_{\s \s}^- &= \frac{1}{\sqrt{2}} f_{\text{bdy}} \Big(\hf;\xi\Big) + \frac{1}{16384 \sqrt{2}} f\Big(\frac{9}{2};\xi\Big) + \frac{1}{327680 \sqrt{2}} f_{\text{bdy}}\Big(\frac{13}{2};\xi\Big)  + \ldots
\end{split}
\ee
The constant term in the $+$ case corresponds to a one-point function of $\sigma$ and therefore the $\mathbb Z_2$ symmetry is broken by the boundary conditions. We can thus identify it with the extraordinary transition. As an additional check one may verify that the bulk block decomposition agrees with the decompositions \eqref{cardystatesord} and \eqref{cardystatesextr}.
 
For completeness, we present the conformal block decomposition for the energy two-point function. We have \cite{DiFrancesco:1997nk},
\be 
G^{\pm}_{\veps \veps} = \xi + \frac{1}{\xi + 1}\, ,
\ee
so this expression is valid for both boundary conditions.
The decomposition in the bulk channel is,
\be 
G^{\pm}_{\veps \veps}  = 1 + \sum^{\infty}_{n=1} \binom{2n-3}{n-2}^{-1} f_{\text{bulk}}(2n;\xi)\, .
\ee
For the boundary expansion we obtain,
\be 
\xi^{-\Delta_{\veps}} G^{\pm}_{\veps \veps}  = 1+\sum^{\infty}_{n=1} \binom{4n-3}{2n-2}^{-1} f_{\text{bdy}}(2n;\xi).
\ee
From the expressions above we learn that the coefficients of the conformal blocks are positive both in the boundary and in the bulk channels.

\subsection{The unitarity minimal models and their analytic continuation}
\label{sec:minimalmods}

Let us now generalize the results of the previous subsection to the whole  series of the unitarity minimal models.
Primary operators in the $(m, m+1)$ model, $m \geq 3$,  
are labeled by integers $(r,s)$, with $1 \leq r \leq m-1$, $1 \leq s \leq m$ and the identification
$(r,s) \sim (m-r, m+1 - s)$.
 Denoting the $(1,2)$ operator by $\sigma$ and the $(1,3)$ operator by $\veps$,  the relevant OPE and scaling dimensions are
\be
\label{minmodope}
\sigma \times \sigma = \mathbf 1 + \veps\,, \qquad \qquad \Delta_{\s} = \hf - \frac{3}{2(m+1)}\,, \qquad \qquad \Delta_{\veps} = 2 - \frac{4}{m+1}\,.
\ee
We can eliminate $m$ to find
\be
\label{deltaeps}
\Delta_{\veps} = \frac{2}{3} (4 \Delta_\s + 1)\, ,
\ee
and we will work with $\Delta_\s$ rather than $m$ as our independent variable from now on.

We are after the $\langle \sigma \sigma \rangle$ correlator with  the Cardy boundary condition labelled by the identity.
(Recall that in the Ising model this Cardy state is associated to the extraordinary transition, see equation (\ref{cardystatesextr})).
This correlator can be obtained as a special case of a result obtained
 in the context of Liouville theory with ZZ boundary conditions
 \cite{Zamolodchikov:2001ah}, where the two-point function
\be
\vev{V_{-b/2} (x) V_\alpha(y)}
\ee
was evaluated. Here $V_\alpha(x)$ denotes the usual Liouville vertex operator with scaling dimension $\Delta_\alpha = \alpha (Q - \alpha)$ with $Q = b + b^{-1}$. We will be interested in the case $\alpha = - b/2$ and $b$ set to the minimal model values given by
\be
c = 1 + 6 Q^2  =  1 - \frac{6}{m (m+1)}\,.
\ee
One may verify that solving this equation for $b$ results in a scaling dimension of $V_{-b/2}$ which is precisely $\Delta_\sigma$ given in \eqref{minmodope}.

The two-point function from \cite{Zamolodchikov:2001ah} takes the form:
\be
\label{minmod2pt}
\begin{split}
G_{\s \s} (\xi) &= 2 \sin\left( \frac{\pi}{6} (1 + 4 \D_\s) \right) \xi^{(4 \D_\s + 1)/3} (1+\xi)^{- (\D_\s + 1)/3} \\ &\qquad \times {}_2 F_1\left(\frac{1 - 2 \D_\s}{3},\frac{2+ 2\D_\s}{3};\frac{2 - 4 \D_\s}{3};\frac{1}{\xi + 1}\right)\,.
\end{split}
\ee
The boundary conformal block decomposition of this correlation functions contains operators with even dimensions,
\begin{multline}
\xi^{-\Delta_\s} G_{\s \s} (\xi) = 2 \sin\left( \frac{\pi}{6} (1 + 4 \Delta_\s) \right)  \Big( 1 + \frac{\Delta_\s  (1+\Delta_\s )}{2(5-4 \Delta_\s )} f_\text{bdy}(2,\xi) + \frac{\Delta_\s  (1+\Delta_\s )^2 (2+5 \Delta_\s )}{40 (11 - 4 \Delta_\s ) (5 - 4 \Delta_\s)}f_{\text{bdy}} (4,\xi)  
\\  \qquad + \frac{\Delta_\s  (1+\Delta_\s )^2 \left(20+106 \Delta_\s +35 \Delta_\s ^2+21 \Delta_\s ^3\right)}{1008 (17- 4 \Delta_\s ) (11- 4 \Delta_\s ) (5-4 \Delta_\s )}f_\text{bdy} (6,\xi) + \ldots \Big)\,,
\end{multline}
in agreement with the fact that the only boundary block is the identity Virasoro block. In the bulk channel we find the identity and the $\veps$ Virasoro blocks, leading to a decomposition into $SO(2,1)$ blocks with operators of dimension of $1 + 4 n$ and $\D_\veps + 4 n$ with $n$ a non-negative integer. For the first few coefficients we find
\begin{align}
&G_{\s \s} (\xi) =  1 + \frac{\Delta_\s  (1+\Delta_\s )}{2(5-4 \Delta_\s )} f_\text{bulk}(4,\xi) + \frac{\Delta_\s  (1+\Delta_\s )^2 (2+5 \Delta_\s )}{40 (11 - 4 \Delta_\s ) (5 - 4 \Delta_\s)}f_{\text{bulk}} (8,\xi)  + \ldots\\
&-\frac{\G\left(\frac{2 - 4 \D_\s}{3}\right) \G\left(\frac{2 + 2\Delta_\s}{3}\right)}{\G(-2 \Delta_\s ) \G\left(\frac{4 + 4\Delta_\s}{3}\right)} \left( f_{\text{bulk}} (\D_\veps,\xi) + \frac{(1+\Delta_\s ) (2+5 \Delta_\s ) (-1+8 \Delta_\s )}{6 (7+4 \Delta_\s ) (5+8 \Delta_\s )} f_{\text{bulk}} (\D_{\veps} + 4,\xi) + \ldots \right) \,.\nonumber
\end{align}
Up to the normalization factor $2 \sin (\frac{\pi}{6} (1 + 4 \D_\s))$, the coefficients of the first series are the same as those of the boundary identity Virasoro block. Indeed, in either channel these blocks correspond to Virasoro descendants of an identity operator. Notice also that the coefficient of the block with dimension $\D_\veps + 4$  has a zero precisely when $\D_\s = \frac{1}{8}$, reflecting the aforementioned fact that $L_{-2} \veps$ is actually an $SO(2,2)$ descendant in the two-dimensional Ising model. (Indeed in the Ising model the $(1,3)$ primary
is identified with the $(2,1)$ primary which has a level-two null descendant.)

Remarkably, the coefficients of the boundary conformal blocks turn out to be positive for $0 < \Delta_\s < \frac{5}{4}$.\footnote{We have verified this statement to high order and believe that it is generally true although we currently cannot offer a rigorous proof.} This implies that we have found a solution of the crossing symmetry equation that is consistent with the unitarity requirements for \emph{any} value of $\Delta_\s$ in this interval, given simply by the analytic continuation of \eqref{minmod2pt} away from the minimal model values for $\D_\s$.  Of course this does not imply that this correlator can always be embedded in a full-fledged unitary CFT -- in fact we already know that this is only possible if $\Delta_\s$ has one of the minimal model values.

As we pointed out repeatedly in this paper, unitarity does not require the coefficients of the bulk channel conformal blocks to be positive. These coefficients however do turn out to be positive for the smaller range $ \frac{1}{8} < \Delta_\s  < 1$. The lower and upper endpoint of this range are determined by the zeroes of the blocks of dimension $\D_\veps + 4$ and $\D_\veps$, respectively.

In summary, for the range $0 < \Delta_\s < \frac{5}{4}$  we have found an exact solution to the boundary crossing symmetry equation (\ref{crossingsy}),
with the dimension of the first bulk scalar primary  $\epsilon$ in the $\sigma \sigma$ OPE given by (\ref{deltaeps}). In the smaller range  $\frac{1}{8} < \Delta_\s  < 1$
 the bulk expansion satisfies positivity.

\subsubsection*{An aside: the four-point function}
The result just found compels us to briefly consider
the analogous analytic continuation
 for the mimimal-model four-point function (without boundary). The numerical bounds in that case \cite{Rychkov:2009ij} appear to converge to a shape with a ``kink'' at the Ising model, so at $\Delta_\sigma = \frac{1}{8}$, which is followed by a straight line that is approximately given by \ref{deltaeps} for $\frac{1}{8} < \Delta_\sigma \lesssim 0.4$. (For larger values of $\Delta_\sigma$ the numerical analysis becomes less precise.)

Now, as we show momentarily, one can easily repeat the boundary analysis and construct a four-point function $\langle \sigma \sigma \sigma \sigma \rangle$
with $\Delta_\veps$ given by \ref{deltaeps}. This solution appears to precisely saturate the numerical bounds of  \cite{Rychkov:2009ij} in the range $\frac{1}{8} < \Delta_\s < \frac{1}{2}$,
which explains why they cannot be lowered by {\it e.g.} improving the numerical accuracy, even in between the discrete minimal model points. Note of course that there is no complete unitary CFT 
when one does not precisely sit at the minimal model points, only a solution for this particular four-point function.
  On the other hand, we currently cannot explain why this solution is ``extremal'' in the sense that it saturates the  bounds.

We can construct this ``interpolating'' solution by noticing that the $\sigma$ field in the minimal models has a null descendant at level two and its correlation functions therefore satisfy the following differential equation: 
\be
\Big(\mathcal L_{-2} - \frac{3}{2 (2h +1)} \mathcal L_{-1}^2\Big) \vev{\sigma(z) \op_1(z_1) \ldots \op_n (z_n) } = 0\,,
\ee
with
\be
\begin{split}
\mathcal L_{-1} &= \del_z\,,\\
\mathcal L_{-2} &= \sum_{i=1}^n \left( \frac{1}{(z- z_i)} \del_{z_i} + \frac{h_i}{(z - z_i)^2}\right)\,.
\end{split}
\ee
Our natural conjecture is that the same differential equation is satisfied by a putative four-point function that saturates the bound and interpolates between the minimal models.

When we apply this differential operator to the four-point function of four $\sigma$ fields, which we write as usual as
\be
\vev{\sigma(x_1)\sigma(x_2)\sigma(x_3)\sigma(x_4)} = \frac{1}{(x_{12})^{2 \Delta} (x_{34})^{2 \Delta}} G(z,\bar z)\,,
\ee
we find a simple hypergeometric equation that we can easily solve. Combining the holomorphic and antiholomorphic part the two solutions become:
\be
\begin{split}
G(z,\bar z) &= G_1(z) G_1(\bar z) + N(\D) G_2(z) G_2(\bar z)\,,\\
G_1(z) &= (1-z)^{-\Delta } {}_2 F_1\left(\frac{1 - 2 \Delta}{3},-2 \Delta ,\frac{2(1 - 2 \Delta)}{3},z\right) = 1 + O(z^2) \,,\\
G_2(z) &= (1-z)^{\frac{1+\Delta }{3}} z^{\frac{1}{3}+\frac{4 \Delta }{3}} {}_2 F_1 \left(\frac{2 (1+\Delta )}{3},1+2 \Delta ,\frac{4 (1+\Delta )}{3},z\right) = z^{(4 \Delta + 1)/3} (1 + O(z)) \,.
\end{split}
\ee
Crossing symmetry fixes the relative normalization to be:
\be
N(\D) = \frac{2^{1-\frac{8 (1+\Delta )}{3}} \Gamma\left(\frac{2}{3}-\frac{4 \Delta }{3}\right)^2 \Gamma(1+2 \Delta )^2 \left(-\cos\left(\frac{1}{3} (\pi +4 \pi  \Delta )\right)+\sin\left(\frac{1}{6} (\pi +16 \pi  \Delta )\right)\right)}{\pi  \Gamma\left(\frac{7}{6}+\frac{2 \Delta }{3}\right)^2}\,.
\ee
Notice that $N(\Delta) > 0$ for $0 < \Delta < 1$ but $N(0) = N(1) = 0$.

The conformal block decomposition can be found most easily by first decomposing the holomorphic functions separately in terms of the holomorphic building blocks
\be
f(\beta,z) = z^\beta {}_2 F_1 (\beta,\beta,2\beta,z)\,.
\ee
For the identity block we then find that
\be
G_1(z) = \sum_{n=0}^\infty c_{1}(n) f(2n,z)\,,\\
\ee
with the first few coefficients given by:
\be
\begin{split}
c_1(0) &= 1\,,\\
c_1(1) &=  \frac{\Delta  (1+\Delta )}{2 (5-4 \Delta )}\,,\\
c_1(2) &= \frac{\Delta  (1+\Delta )^2 \left(1+\frac{5 \Delta }{2}\right)}{20 (-11+4 \Delta ) (-5+4 \Delta )}\,,\\
c_1(3) &= -\frac{\Delta  (1+\Delta )^2 \left(5+\frac{1}{2} \Delta  \left(53+\frac{7}{2} \Delta  (5+3 \Delta )\right)\right)}{252 (-17+4 \Delta ) (-11+4 \Delta ) (-5+4 \Delta )}\,,
\end{split}
\ee
and we checked that the first six coefficients are all positive functions for $0 < \Delta < 1$.

The epsilon block can be decomposed as:
\be
G_2(z) = \sum_{n=0}^\infty c_2(n) f\left(\frac{4 \Delta + 1}{3} + 2n,z\right)\,,
\ee
with
\be
\begin{split}
c_2(0) &= 1\,,\\
c_2(1) &= \frac{(1+\Delta ) \left(1+\frac{5 \Delta }{2}\right) (-1+8 \Delta )}{3 (7+4 \Delta ) (5+8 \Delta )}\,,\\
c_2(2) &= \frac{(1+\Delta )^2 \left(-35+\frac{1}{2} \Delta  \left(419+\frac{1}{2} \Delta  (6315+64 \Delta  (97+25 \Delta ))\right)\right) }{18 (7+4 \Delta ) (13+4 \Delta ) (11+8 \Delta ) (17+8 \Delta )} \,. 
\end{split}
\ee 
We again checked by hand that the first eight coefficients are positive for $\frac{1}{8} < \Delta < 1$. They however also all have simple zeroes for some $\Delta \leq \frac{1}{8}$, starting with $c_2(1)$ at $\Delta = \frac{1}{8}$. This first zero corresponds to the decoupling of a spin 2, dimension 3 operator (as well as an infinite number of its Virasoro descendants) which we can again identify with the level two null descendant of the epsilon operator in the two-dimensional Ising model. The fact that our putative solution extends to $\Delta = 1$ implies that it extrapolates beyond the accumulation points of the minimal models at $\Delta = \frac{1}{2}$.

Notice that for $\Delta_\s > \frac{1}{2}$ it is natural to conjecture that the bound will be saturated instead by the free boson where (in the notation of this section) $\D_\eps = 4 \Delta_\sigma$. On the other hand, for $\Delta_\s < \frac{1}{8}$ our putative solution becomes invalid precisely because the coefficient of the conformal block corresponding to $L_{-2} \veps$ becomes negative. It would be very interesting to find a solution saturating the numerical bound for $\Delta_\s < \frac{1}{8}$ which would would allow one to investigate the precise transition at the ``corner'' corresponding to the Ising model point and the role of $L_{-2} \veps$ in this transition. Interestingly, the (approximate) decoupling of the first irrelevant spin 2 operator was observed numerically in three dimensions as well \cite{ElShowk:2012ht}.

\subsection{$\langle \phi^2 \phi^2 \rangle$ correlator}

In this section we will decompose $\langle \phi^2 \phi^2 \rangle$ in free field theory. This expansion complements the order $\epsilon$ expression for the scalar two-point function of section \ref{sec:bootstrapinepsexpansion}. 
The $\phi^2$ two-point function is,
\be
\begin{split}
G^{\pm}_{\phi^2\phi^2} &= \left(1 \pm \left(\frac{\xi}{\xi + 1}\right)^{\hf d-1}\right)^2 + \frac{N}{2}\xi^{d-2}\, ,
\end{split}
\ee 
where the plus/minus sign corresponds to Neumann/Dirichlet boundary conditions, and $N$ is the number of scalars.
The conformal block expansion in the bulk channel is
\be
G^{\pm}_{\phi^2\phi^2} = 1 + \l a_{\phi^2}f_{\text{bulk}}(d-2;\xi) + \sum_{n=0}^{\infty}\l a_{\phi^4, n} f_{\text{bulk}}(2d-4+2n;\xi)\, ,
\ee 
with
\be
\begin{split}
\l a_{\phi^2}    & = \pm 2\, ,\\
\l a_{\phi^4, n} & = \frac{\left((-1)^n 2^d \Gamma(\frac{d-1}{2})\Gamma(\hf d+n-1)+4 N \sqrt{\pi}\Gamma(d+n-2)\right)\Gamma(d+n-2)\Gamma(\tf d+n-4)}
{8\sqrt{\pi}\Gamma(d-2)^2\Gamma(n+1)\Gamma(\tf d+2n-4)}\, .
\end{split}
\ee 
The Neumann expansion exhibits positivity while the Dirichlet case has one negative coefficient.
In the boundary channel we have,
\be
\xi^{-d+2}G^{\pm}_{\phi^2\phi^2} = \frac{N}{2} + \sum_{n=0}^{\infty}\mu^2_n f_\text{bdy} (d-2+2n; \xi)\, ,
\ee
with
\be 
\mu^2_n=(1 \pm \delta_{n,0})\frac{4^{1-n}}{(2n)!}\frac{\Gamma(\frac{d-1}{2}+n)\Gamma(\hf d+n-1)\Gamma(d+2n-3)}{\Gamma(\hf d-1)\Gamma(d-2)\Gamma(\frac{d+4n-3}{2})}\, ,
\ee
with positivity in both cases. 

\subsection{The extraordinary transition}
\label{subsec:phiextraord}
There is no extraordinary transition in $4$ dimensions since the conformally invariant one-point function of a free field is not compatible with its equation of motion. In $4-\epsilon$ dimensions the equation of motion is however modified to:
\be
\square \phi = \frac{\lambda_*}{6} \phi^3
\ee
with $\lambda_* = 48 \pi^2 \epsilon / (N+8)$ and $N=1$ in our case. On the half-space this equation admits the solution:
\be
\label{vevphi}
\vev{\phi(x)} = \sqrt{\frac{12}{\lambda_*}} \frac{1}{x^d}
\ee
which to leading order is consistent with boundary conformal invariance. This solution is our starting point for the analysis of the extraordinary transition in the $\epsilon$ expansion. 

Let us compute the two-point function of the scalar field $\phi$. We may shift the field $\phi$ by its classical one-point function,
\be
\phi(x) = \vev{\phi(x)} + \chi(x)
\ee
and find the propagator $G(x,y) = \vev{\chi(x) \chi(y)}$ by solving the linearized equation of motion around this solution,
\be
\left(\square - \frac{6}{(x^d)^2} \right) G(x,y) = \delta^d(x-y)
\ee
The solution compatible with the boundary conditions at $x^d = 0$ takes the form:
\be
\label{propchi}
\begin{split}
G(x,y) &= \frac{1}{(2 x^d)(2 y^d)} \xi^{-1} \left( \frac{1}{4\pi^2} G^0(\xi) \right)\\
G^0(\xi) &=  \frac{1}{1 + \xi} +12\xi  +6 \xi   (1+2 \xi ) \log\left(\frac{\xi}{1+\xi}\right)
\end{split}
\ee
with $\xi = (x-y)^2 / (4 x^d y^d)$, as before. On the first line we recognize the familiar form of a scalar two-point function for a CFT with a boundary. Taking the limit $\xi \to 0$ in \eqref{propchi} we see that the properly normalized operator is actually $2 \pi \chi$  rather than $\chi$, and similarly $2 \pi \phi$ rather than $\phi$. We will henceforth work with these rescaled operators. This implies that from now on $\vev{\phi(x)} = 3/(\sqrt{\epsilon}\, x^d)$ and we can drop the $4 \pi^2$ on the first line of \eqref{propchi}.
 
We will now expand the two-point function of $\phi$ in conformal blocks. It is important to note that OPE statements always refer to full correlation functions, \ie including any disconnected contributions. In our case the disconnected part $\vev{\phi(x)}\vev{\phi(y)}$ is of order $1/\eps$ which makes it the leading-order term. Our first task is thus to decompose $\vev{\phi(x)}\vev{\phi(y)}$ in conformal blocks. In the boundary channel we of course find precisely the block corresponding to the identity operator and nothing else. In the bulk channel we find:
\be
\label{blockdecdisc}
\vev{\phi(x)}\vev{\phi(y)} = \frac{36 / \e}{(2 x^d) (2 y^d)} =  \frac{36 / \e}{(2 x^d) (2 y^d)} \xi^{-1} \left(f_{\text{bulk}}(2,\xi) + \sum_{n=1}^{\infty} \frac{2 (n!)^2}{(2n)!} f_{\text{bulk}}(2 + 2n,\xi)\right)
\ee
Interestingly, the product of two one-point functions decomposes into an infinite set of bulk blocks with dimensions given by the even integers and with positive coefficients. However, as expected for a totally disconnected correlator, the \emph{bulk} identity operator is missing at this order.

At the next order we should take into account that the one-point function of $\phi$ a priori has subleading corrections,
\be
\vev{\phi(x)} =  \frac{3}{\sqrt \eps\,x^d} \left( 1 + \epsilon\, a + \frac{\eps}{2} \log(2 x^d) \right)
\ee 
with an unknown coefficient  $a$ and with the logarithm originating from the correction to the scaling dimension of $\phi$ in $4-\eps$ dimensions. The full two-point function to order $\epsilon^0$ becomes:
\be
\begin{split}
\vev{\phi(x)\phi(y)} &= \frac{1}{(2 x^d)^{\Delta_\phi} (2 y^d)^{\Delta_\phi}} \xi^{-\Delta_\phi} G^{\text{ext}}(\xi)\\
G^{\text{ext}}(\xi) &= \frac{36}{\eps}(1 + 2 \eps\, a) \xi - 18 \xi \log(\xi) + G^0(\xi)
\end{split}
\ee
with $\Delta_\phi = 1 - \eps/2$ the free-field dimension of $\phi$ in $4-\epsilon$ dimensions.

In the boundary channel the conformal block decomposition of this corrected correlator is again straightforward. The corrections to the disconnected part of course simply become corrections to the boundary identity block, whilst for the connected part we find that:
\be
\label{epsextrabdy}
\xi^{-1} G^0(\xi) = \frac{1}{10} f_\text{bdy}(4;\xi)
\ee
so we find a single boundary block of dimension $d = 4$. This is completely as expected. In particular, the existence of a gap of size $d$ was an essential assumption in the numerical bootstrap for the bulk bounds.

In the bulk channel we find subleading corrections to the infinite series of blocks in \eqref{blockdecdisc} but no new blocks. The first few terms take the form:
\be
\label{blockdecextr}
\begin{split}
G^{\text{ext}}(\xi) &= 1 + \left(\frac{36}{\eps} + 11 + 72 a\right) f_\text{bulk}\left(2 - \frac{2}{3}\eps;\xi\right) \\
&\qquad + \left(\frac{36}{\eps} - 12 + 72 a\right) f_\text{bulk}\left(4 ,\xi\right) + \left(\frac{12}{\eps} - 18 + 24 a\right) f_\text{bulk}(6 + 2 \eps, \xi) \\ &\qquad+ \left(\frac{18}{5 \eps} + \frac{1}{20}(-241 + 144 a)\right) f_\text{bulk}\left(8 + \frac{16}{3} \eps,\xi\right) + \ldots
\end{split}   
\ee
where it is understood that the blocks are evaluated in $4-\epsilon$ spacetime dimensions. The order $1/\epsilon$ terms in \eqref{blockdecextr} of course coincide with \eqref{blockdecdisc}. The identity operator is now present in the bulk channel, and the dimension of the next operator (which is $2 - \frac{2}{3}\eps$) is precisely the one-loop dimension of $\phi^2$ in the epsilon expansion. It would be interesting to compute $a$ so we can get an idea of positivity of the coefficients for $\eps = 1$.

\subsection{A trivial solution}
\label{subsec:trivialsol}
A particularly simple solution of \eqref{crossingsy} is obtained by assuming that the bulk channel only contains the identity operator, so all the non-trivial one-point functions are set to zero. In that case there is effectively no boundary at all and the two-point function is just $(x_1-x_2)^{-2\Delta}$. This two-point function still has a boundary conformal block decomposition of the form:
\be
\xi^{-\Delta}  = \sum_{m=0}^{\infty} \mu_m^2 f_{\text{bdy}}(\D + m;\xi)  
\ee 
with
\be
\mu_m^2 =
\begin{cases}
\frac{1}{2^{m} m!}(\Delta)_m (\Delta - \frac{d}{2} + 1)_{m/2} (\Delta - \frac{d-1}{2} + m)_{-m/2} &  m \text{ even}\\
\frac{1}{2^m m!} (\Delta)_m (\Delta - \frac{d}{2} + 1)_{(m - 1)/2} (\Delta - \frac{d-1}{2} + m)_{(1-m)/2} & m \text{ odd}
\end{cases} 
\ee
All the coefficients are positive for $\Delta$ greater than the unitarity bound and the boundary spectrum begins with an operator of dimension $\Delta$.

\subsection{Generalized free field theory}
As a simple generalization of the free field theory result we define generalized free field (or gff) two-point functions in the presence of a boundary as:
\be
\begin{split}
\vev{O(x_1) O(x_2)} &= \frac{1}{(x_1- x_2)^{2 \Delta}} \pm \frac{1}{\big( (x_1-x_2)^2 + 4 x^d_1 x^d_2 \big)^\Delta} \\ &= \frac{1}{(2x^d_1)^\Delta (2 x^d_2)^\Delta} \xi^{-\Delta} G^{\pm}_{\text{gff}} (\xi) \qquad\qquad G^{\pm}_{\text{gff}} (\xi) = 1 \pm \left(\frac{\xi}{\xi + 1} \right)^{\Delta}
\end{split}
\ee  
The conformal block decomposition in the bulk takes the form
\be
G^{\pm}_{\text{gff}}(\xi) = 1 \pm \sum_{n=0}^\infty \frac{(-1)^n (\Delta)_n \left(2\D - \frac{d}{2} + 2n\right)_{-n}}{\left(\D - \frac{d}{2} + n + 1\right)_{-n} n!} f_{\text{bulk}} (2 \Delta + 2n ; \xi)
\ee
which has the expected `double trace' infinite operator spectrum and coefficients with alternating signs. On the boundary we find that:
\be
\begin{split}
\xi^{-\Delta} G^{+}_{\text{gff}}(\xi) &= \sum_{n=0}^{\infty}\frac{(\Delta)_{2n} \left(\Delta - \frac{d-1}{2} + 2n \right)_{-n}}{2^{2n-1} (2n)! \left(\Delta - \frac{d}{2} + n + 1\right)_{-n}} f_{\text{bdy}}(\D + 2n; \xi)\\
\xi^{-\Delta} G^{-}_{\text{gff}}(\xi) &= \sum_{n=0}^{\infty} \frac{(\Delta)_{2n+1} \left(\Delta - \frac{d- 3}{2} + 2n \right)_{-n}}{2^{2n}(2n+1)! \left(\Delta - \frac{d}{2} + n + 1\right)_{-n}} f_{\text{bdy}} (\D + 2n + 1; \xi)
\end{split}
\ee
and we find two `single trace' operator spectra on the boundary, both with positive coefficients.

\subsection{$O(N)$ model at large $N$}
 \label{sec:ONlargeN}
For the the $O(N)$ model with Neumann boundary conditions the scalar two-point function is given by \cite{McAvity:1995zd},
\be 
G_{O(N)} = \left(\frac{1}{1+\xi}\right)^{\hf d-1}(1+2\xi)\, .
\ee
The bulk channel expansion is,
\be 
G_{O(N)} = 1+\sum^{\infty}_{n=0} \lambda a_n f_{\text{bulk}}(2n+2;\xi)\, ,
\ee
with
\be 
\lambda a_n = (-1)^{2n}\frac{(d^2-4d(n+2)+8(1+n)^2+4)\Gamma{(1-\hf d+n)\Gamma{(2-\hf d+n)}^2}}{4\Gamma{(2-\hf d)}^2\Gamma(n+2)\Gamma(2-\hf d + 2n)}\, .
\ee
As in all the expansions with Neumann boundary conditions studied in this appendix, the bulk channel coefficients are positive. Finally, the boundary channel expansion is,
\be 
\xi^{-\hf d+1}G_{O(N)} = 2f_{\text{bdy}}(d-3;\xi)\, .
\ee
It is somewhat unexpected that in this channel we have a single block.

\section{Conformal block decompositions for $T_{\m \n}$}
\label{app:enmomtencorrs}
In this appendix we present a few explicit examples of conformal block decompositions of the form \eqref{crossingsyTmn} for the two-point function of the stress tensor.

\subsection{Two bulk dimensions}
The conformal block decomposition of the stress tensor two-point function in two dimensions is a bit subtle, see \cite{McAvity:1993ue} for details. First of all, the residual Virasoro symmetry plus the absence of energy flow across the boundary completely determines the two-point function. Furthermore, the number of independent tensor structures decreases to two and the two functions $f(\xi)$ and $g(\xi)$ have to be replaced with the single function $2 \xi f(\xi) + (1 + \xi) g(\xi)$. With our unit normalization we find that the resulting two-point function is given precisely by the boundary scalar block, with a coefficient that is equal to $4$. In the bulk channel we find the identity plus a single block of dimension $4$ with unit coefficient.

\subsection{Free field theory for general $d$}
The two-point function of the stress tensor in free field theory for $d > 2$ decomposes into infinitely many blocks in either channel. Without presenting all the formulas, we have presented the first few operators and their associated coefficients in both the bulk and the boundary channel in the tables. Notice that the coefficients in the bulk channel are not positive for either boundary condition. 

\begin{table}[h!]
\centering
\begin{tabular}{c|c}
$\D$ & $\lambda a_{\op}$\\
\hline \hline
$0$ & 1 \\ 
$d- 2$ & $\pm \left(\frac{(-2+d) d (1+d)}{4 (-1+d)} \right)$\\
$2d$ & $+ 1$ \\
$2d + 2$ & $- \left(\frac{(-2+d) d}{ (2+3 d)}\right)$\\
$2d+4$ & $+\left(\frac{(-2+d) d^2 (1+d)}{6  (2+d) (4+3 d)}\right)$\\
$2d+6$ & $-\left(\frac{(-2+d) d^2 (1+d) (2+d)}{18 (8+3 d) (10+3 d)}\right)$\\
$2d + 2m$ & $\ldots$
\end{tabular}
\caption{\label{tab:Tmnblockdecbulk}Bulk conformal block decomposition of the two-point function of the stress tensor in free field theory. The first block corresponds to the identity operator and its coefficient sets the overall normalization. The plus/minus sign corresponds to the special/ordinary transition, i.e. Neumann/Dirichlet boundary conditions.}
\end{table}

\begin{table}[h!]
\centering
\begin{tabular}{c|c|c}
$\D$ & $l$ & $\mu^2$\\
\hline \hline
$d$ & $0$ & $\frac{2 d}{(-1+d)}$\\
$d$ & $2$ &  $2^{1-2 d} (1\pm 1)$\\
$d+ 2$ & $2$ & $\frac{2^{-2-2 d} d (-1 + d) (2+d)}{(1+d)}$\\
$d+ 4$ & $2$ & $\frac{2^{-6-2 d} d (-1 + d) 2+d)^2 (4+d)}{3 (7+d)}$\\
$d+2m$ & $2$ & \ldots
\end{tabular}
\caption{\label{tab:Tmnblockdecbdy}Boundary conformal block decomposition of the two-point function of the stress tensor in free field theory.}
\end{table}

\subsection{Extraordinary transition}
In this subsection we compute the two-point function of the stress tensor in the extraordinary transition.
 
The classical stress tensor for the $\lambda \phi^4$ theory with a curvature coupling $z$ takes the form:
\be
\label{Tmnscalar}
T_{\m \n}(x) = \frac{2}{\sqrt{3}} \left( \left(2 z - \hf \right) \del_\m \phi \del_\n \phi  + 2 z\, \phi \del_\m \del_\n \phi + g_{\m \n} \left( \frac{\lambda}{48} \phi^4 + \left(\qt - 2 z\right) \del_\r \phi \del^\r \phi - 2 z \phi \square \phi\right) \right)
\ee
One may easily verify that it is traceless in $d=4$ for $z = 1/12$ which therefore corresponds to the conformally coupled scalar. We will henceforth use $z = 1/12$. In that case $T_{\m \n}$ is unit normalized in free field theory, more precisely $\vev{T_{\m \n} (x) T_{\r \s}(y)} = \frac{H_{12}^2}{(4 \xi)^4 (x^d)^4 (y^d)^4}$ provided $\vev{\phi(x) \phi(y)} = \frac{1}{(x-y)^{2}}$.

The correlation functions of the scalar $\phi$ were computed to leading order in subsection \ref{subsec:phiextraord}. Upon substituting the solution $\vev{\phi(x)} = 3 / (\sqrt \e \, x^d)$ in \eqref{Tmnscalar} we find that the one-point function of $T_{\m \n}$ vanishes, in agreement with the requirements of boundary conformal invariance. At the next order we substitute $\phi(x) = \vev{\phi(x)} + \chi(x)$ and expand in $\eps$ to find an expression of the form:
\be
\label{expansionTmn}
T_{\m \n}(x) = \frac{1}{\sqrt{\eps}} {\cal T}_{\m \n}[x^d, \del_x] \chi(x) + \ldots
\ee
where ${\cal T}_{\m \n}[x^d,\del_x]$ is a linear differential operator which explicitly depends on $x^d$. To leading order we therefore obtain that
\be
\vev{T_{\m \n}(x) T_{\r \s}(y)} = \frac{1}{\e} {\cal T}_{\m \n}[x^d,\del_x] {\cal T}_{\r \s}[y^d,\del_y] \vev{\chi(x)\chi(y)}
\ee
We can now substitute the solution $G(x,y) = \vev{\chi(x) \chi(y)}$, which is equation \eqref{propchi} without the factor of $4 \pi^2$, work out the action of the differential operators $\cal T$ and collect various terms to eventually find a two-point function of the form:
\be
\label{TTextraord}
\vev{T_{\m \n} (x) T_{\r \s}(y)} =  \, \frac{f^\text{ext} (\xi) H_{12}^2 + g^\text{ext}(\xi) H_{12} Q_{12} + h^\text{ext}(\xi) Q_{12}^2}{(4\xi)^4 (x^d)^4 (y^d)^4} 
\ee 
where the tensor structures $H_{12}$ and $Q_{12}$ are defined (in the projective cone notation) in \eqref{H12W12} and \eqref{Q12} and
\be
\label{fghextraord}
f^{\text{ext}}(\xi) = \frac{16 \xi }{\eps (1+\xi )^3}
\qquad  g^{\text{ext}}(\xi) = \frac{64 \xi ^2 (2+5 \xi )}{\eps (1+\xi )^4} \qquad 
h^{\text{ext}}(\xi) = \frac{64 \xi ^3 \left(1+5 \xi +10 \xi ^2\right)}{\eps  (1+\xi )^5}
\ee
Upon comparing \eqref{TTextraord} with the last equation in \eqref{tensorcorrelatorsbdycft} we see that this correlation function has exactly the right tensor structure to be consistent with boundary conformal invariance. Furthermore, the functions $(f^\text{ext},g^\text{ext},h^\text{ext})$ also satisfy the Ward identities \eqref{wardidsbdycorr}. These are rather non-trivial checks of our result.

The conformal block decomposition of \eqref{TTextraord} turns out to be remarkably simple. In the boundary we find only a scalar block (which must have dimension $d$ by the Ward identities) with coefficient $ 640 / \epsilon$. In the bulk we find three blocks,
\be
\label{bulkblockdecTmnextraord}
\begin{pmatrix}
f^\text{ext}(\xi) \\ g^\text{ext}(\xi) \\ h^\text{ext} (\xi)
\end{pmatrix}
= \frac{160}{\epsilon}
\begin{pmatrix}
f_\text{bulk}(2;\xi) \\  g_\text{bulk}(2;\xi) \\ h_\text{bulk}(2;\xi)
\end{pmatrix}
+ 
\frac{480}{\eps}
\begin{pmatrix}
f_\text{bulk}(4;\xi) \\ g_\text{bulk}(4;\xi) \\ h_\text{bulk}(4;\xi)
\end{pmatrix} 
+ 
\frac{320}{\eps}
\begin{pmatrix}
f_\text{bulk}(6;\xi) \\ g_\text{bulk}(6;\xi) \\ h_\text{bulk}(6;\xi)
\end{pmatrix} 
\ee
all with positive coefficients. Notice that the identity operator is absent at this order.

Closer inspection of \eqref{bulkblockdecTmnextraord} leads to a subtlety that we would like to clarify. We easily identify the bulk block with dimension $2$ as the operator $\phi^2$. It appears in the $TT$ OPE with an order one coefficient and its one-point function is $\vev{\phi^2} = \vev{\phi}^2 \sim \eps^{-1}$ so altogether it appears at the right order in $\epsilon$. The counting for the operator of dimension $4$ is however a bit different. The only scalar primary of that dimension is $\phi^4$ but its one-point function is of order $\eps^{-2}$. Our result can therefore only be consistent if $\phi^4$ appears in the stress tensor OPE only at order $\eps$. It is in fact easy to see that the leading-order Feynman diagram for the $\vev{T T \phi^4}$ tree-point function (which would be of order $\eps^0$) has to vanish. This is because it factorizes into a product of two Feynman diagrams that each correspond to the $\vev{T \phi^2}$ two-point function, which in turn vanishes by conformal invariance. This is also consistent with the fact that no dimension $4$ block appears in the bulk conformal block decomposition of the stress tensor two-point function in free-field theory, cf. table \ref{tab:Tmnblockdecbulk}. From these tables we may also deduce that a similar cancellation should occur for the dimension $6$ operator.

\bibliographystyle{utphys}
\bibliography{biblio}

\end{document}